\documentclass[sigplan,10pt]{acmart}

\acmConference[Preprint]{Preprint}{2026}{arXiv}
\acmYear{2026}
\setcopyright{none}
\acmISBN{}
\acmDOI{}

\usepackage{booktabs}
\usepackage{array}
\usepackage{subcaption}
\usepackage{listings}
\usepackage{xspace}

\newcommand{\sys}{ForgeKernel\xspace}
\newcommand{\cannbench}{CANN-Bench\xspace}

\newcommand{\SelfEvoTarget}{+5\%}                 % per-phase self-reference target
\newcommand{\goalagent}{\texttt{/goal} agent}  % Claude Fable 5.1 + stock Claude Code CLI in native /goal mode
\newcommand{\SelfEvoDsvFusedInitial}{+5.3\%}
\newcommand{\SelfEvoDsvFusedGenOne}{+10.0\%}
\newcommand{\SelfEvoDsvFusedCumul}{+15.5\%}

\newcommand{\GqaCasePersistentGain}{+0.8\%}

\newcommand{\GqaCaseQSlotsGain}{+6.1\%}

\newcommand{\NumLibraryCells}{22}
\newcommand{\NumLibraryBeat}{18}

\newcommand{\NumLibraryNear}{3}
\newcommand{\SpeedupMeanLibrary}{+1.6\%}

\newcommand{\NumSystemCells}{7}

\newcommand{\SpeedupMeanSystem}{+39.4\%}

\newcommand{\NumOps}{27}
\newcommand{\NumShapes}{29}

\newcommand{\SpeedupMean}{+10\%}
\newcommand{\SpeedupRatioMean}{1.10}
\newcommand{\NumBeat}{25}

\newcommand{\SelfEvoGdnFwdAgentrefGenOne}{+9.5\%}

\newcommand{\SelfEvoBaseGdnFwdAgentref}{+82.6\%}
\newcommand{\SelfEvoCumulGdnFwdAgentref}{+100\%}
\newcommand{\SelfEvoAttnFwdAgentrefGenOne}{+5.0\%}

\newcommand{\SelfEvoBaseAttnFwdAgentref}{+22.8\%}
\newcommand{\SelfEvoCumulAttnFwdAgentref}{+29\%}
\newcommand{\SelfEvoDsaSparseFwdGenOne}{+7.2\%}

\newcommand{\SelfEvoBaseDsaSparseFwd}{$-$0.5\%}
\newcommand{\SelfEvoCumulDsaSparseFwd}{+7\%}
\newcommand{\SelfEvoGdnPrefillBwdGenOne}{+2.0\%}

\newcommand{\SelfEvoBaseGdnPrefillBwd}{+1.1\%}
\newcommand{\SelfEvoCumulGdnPrefillBwd}{+3\%}
\newcommand{\SelfEvoAttnFwdGqaGenZero}{+0.3\%}

\newcommand{\SelfEvoAttnFwdGqaGenOne}{+9.9\%}

\newcommand{\SelfEvoAttnFwdGqaGenTwo}{+4.2\%}

\newcommand{\SelfEvoCumulAttnFwdGqa}{+15\%}

\newcommand{\GainOverAgent}{9.5\%}

\newcommand{\AscendOpCount}{8}
\newcommand{\AscendGeomeanLead}{+25.1\%}

\newcommand{\AscendHw}{910C}
\newcommand{\AscendBenchVersion}{1.1.2}

\newcommand{\LeaderboardSnapshot}{2026-10-30}
\newcommand{\CannSeasonSnapshot}{2026-10-15}
\newcommand{\CannVTwoVersion}{2.0}
\newcommand{\CannVTwoBoards}{Ascend 910C, 950DT, and 950PR}
\newcommand{\CannSeason}{1}
\newcommand{\CannLeaderboardVersion}{1.1}
\newcommand{\CannLeaderboardHw}{910C}
\newcommand{\CannLeaderboardOps}{53}
\newcommand{\CannLeaderboardCases}{1,060}
\newcommand{\CannLeaderboardScore}{8566.1110}
\newcommand{\CannStandardScore}{4362.7547}
\newcommand{\CannHiddenScore}{4203.3563}
\newcommand{\CannHiddenCases}{4,240}
\newcommand{\CannHiddenPassed}{4,185}
\newcommand{\CannStandardSpeedup}{1.759}
\newcommand{\CannHiddenSpeedup}{1.854}
\newcommand{\CannCombinedSpeedup}{1.835}
\newcommand{\SolExecStack}{v1.1}
\newcommand{\SolExecHw}{B200}
\newcommand{\SolExecProblems}{26}
\newcommand{\SolExecScore}{0.815204}
\newcommand{\SolExecWorkloads}{660}
\newcommand{\SolExecFastCount}{647}
\newcommand{\SolExecFastPercent}{98.03\%}
\newcommand{\SolExecAvgSpeedup}{5.54}

\usepackage{pgfplots}
\pgfplotsset{compat=1.18}
\usepgfplotslibrary{groupplots}
\definecolor{fkGain}{HTML}{2A78D6}   % at or above the reference
\definecolor{fkLoss}{HTML}{E34948}   % below the reference
\definecolor{fkRamp1}{HTML}{86B6EF}
\definecolor{fkRamp2}{HTML}{2A78D6}
\definecolor{fkRamp3}{HTML}{104281}
\definecolor{fkInk}{HTML}{52514E}    % axis and label ink
\definecolor{fkGrid}{HTML}{E5E4E0}   % one step off the page
\definecolor{fkMuted}{HTML}{9A9A96}  % pending / de-emphasised marks
\pgfplotsset{
  fkbase/.style={
    scale only axis, axis lines=left, axis line style={fkInk, thin},
    tick align=outside, tick style={fkInk, thin},
    every axis label/.append style={font=\scriptsize, fkInk},
    tick label style={font=\scriptsize, fkInk},
    legend style={font=\scriptsize, draw=none, fill=none, inner sep=1pt,
                  /tikz/every even column/.append style={column sep=5pt}},
    grid style={fkGrid, thin},
    clip=false,
  },
  fkgainbar/.style={
    fkbase, xbar, bar width=4.5pt, y dir=reverse, ytick style={draw=none},
    xmajorgrids, ymajorgrids=false,
    y axis line style={draw=none}, ytick style={draw=none}, enlarge y limits=false,
    xticklabel={\pgfmathprintnumber{\tick}\%},
    every node near coord/.append style={font=\scriptsize, fkInk, inner sep=1.5pt},
    yticklabel style={align=right},
  },
  fkgaindot/.style={
    fkbase, y dir=reverse, only marks, mark size=2.1pt,
    xmajorgrids, ymajorgrids=false, y axis line style={draw=none}, ytick style={draw=none}, enlarge y limits=false,
    xticklabel={\pgfmathprintnumber{\tick}\%},
    every node near coord/.append style={font=\scriptsize, fkInk, inner sep=1.5pt},
    yticklabel style={align=right},
  },
  fkablbar/.style={
    fkbase, ybar, bar width=4.5pt, ymajorgrids, xtick style={draw=none},
    area legend,   % one swatch per configuration, not a pair of stub bars
    x axis line style={draw=none}, enlarge x limits=0.14,
    yticklabel={\pgfmathprintnumber{\tick}\%},
    every node near coord/.append style={font=\tiny, fkInk, inner sep=1pt},
  },
  fkdumbbell/.style={
    fkbase, only marks, mark size=2.3pt, y dir=reverse,
    xmajorgrids, ymajorgrids=false, y axis line style={draw=none}, ytick style={draw=none}, enlarge y limits=false,
  },
}

\newcommand{\affTsinghua}{\affiliation{%
  \institution{Tsinghua University}
  \city{Beijing}
  \country{China}}}
\newcommand{\affModelBest}{\affiliation{%
  \institution{ModelBest Inc.}
  \city{Beijing}
  \country{China}}}
\newcommand{\affUCAS}{\affiliation{%
  \institution{University of Chinese Academy of Sciences}
  \city{Beijing}
  \country{China}}}
\newcommand{\affHIT}{\affiliation{%
  \institution{Harbin Institute of Technology}
  \city{Harbin}
  \country{China}}}

\begin{document}

%% Short form feeds the running head; the \\ break is the ICLR edition's.
\title[ForgeKernel: Self-Evolving Kernels]{ForgeKernel: Self-Evolving Kernels\\
Catching and Surpassing Experts and Agents}

\author{Yaojian Chen}
\authornote{The first three authors contributed equally to this work.}
\affTsinghua
\email{yj-chen21@mails.tsinghua.edu.cn}

\author{Dazhi Peng}
\authornotemark[1]
\affModelBest
\email{pengdazhi@modelbest.cn}

\author{Yiyun Zheng}
\authornotemark[1]
\affModelBest
\email{zhengyiyun@modelbest.cn}

\author{Zhui Zhu}
\affTsinghua
\email{z-zhu22@mails.tsinghua.edu.cn}

\author{Shengye Xiang}
\affTsinghua
\email{xcy23@mails.tsinghua.edu.cn}

\author{Leshan Li}
\affTsinghua
\email{lils25@mails.tsinghua.edu.cn}

\author{Zhen Li}
\affModelBest
\email{lizhen@modelbest.cn}

\author{Xianglong Deng}
\affUCAS
\email{dengxianglong02@gmail.com}

\author{Haojun Sun}
\affModelBest
\email{sunhaojun@modelbest.cn}

\author{Qingfeng He}
\affTsinghua
\email{heqf21@mails.tsinghua.edu.cn}

\author{Shangzhan Li}
\affHIT
\email{szli@ir.hit.edu.cn}

\author{Yifei Shen}
\affModelBest
\email{shenyifei@modelbest.cn}

\author{Lin Gan}
\affTsinghua
\email{lingan@tsinghua.edu.cn}

\author{Yuxuan Li}
\authornote{Corresponding authors.}
\affModelBest
\email{yxuanl1995@gmail.com}

\author{Xu Han}
\authornotemark[2]
\affTsinghua
\email{han-xu@tsinghua.edu.cn}

\author{Zhiyuan Liu}
\authornotemark[2]
\affTsinghua
\email{liuzy@tsinghua.edu.cn}

\begin{abstract}
Purpose-built kernel agents can now write high-performance GPU kernels. As
base models grow stronger, though, even a naive agentic loop beats human
experts on many operators, so both the purpose of a kernel agent and the way
it should be designed need rethinking. Yet existing kernel agents generally
lack cross-run knowledge accumulation and reuse, as well as continued
evolution at the architectural level: each run starts from the same initial
knowledge and merely outputs the best-performing kernel found during that
search, without retaining the knowledge gained through optimization for
subsequent runs; meanwhile, once the kernel architecture is fixed early,
later iterations are largely confined to local optimization within it.

We present \sys, a kernel-generation system for continued optimization. It is
designed to accumulate and reuse knowledge across runs while enabling continued
evolution of kernel architectures. It combines two mechanisms. First, it
analyzes the hardware behavior of a high-performance reference and distills
its validated architectural decisions into reusable knowledge of hardware
mechanisms rather than imitating its source code. Second, it self-evolves by
sealing a kernel that passes correctness and performance validation and using
it as the reference for the next optimization run, allowing the optimization
results of one run to inform subsequent runs and providing a basis for
revisiting and adjusting the kernel architecture.

Across \NumOps{} operators and \NumShapes{} input configurations on NVIDIA Hopper
and Blackwell GPUs, \sys matches or beats its expert reference on \NumBeat{} of
them, for a \SpeedupMean{} geometric mean. On Ascend \AscendHw{}, on the public case set of all \AscendOpCount{}
level-4 operators in the \cannbench{} Season~\CannSeason{} final leaderboard,
\sys leads the per-operator public best, with a \AscendGeomeanLead{}
geometric-mean lead, while its full-suite submission ranks first on that
benchmark's A3 leaderboard. As of \LeaderboardSnapshot{}, public submissions
generated by \sys rank first on all three \cannbench{}~\CannVTwoVersion{}
solution leaderboards (\CannVTwoBoards{}) and on the SOL-ExecBench
FlashInfer-Bench track.

\end{abstract}


\maketitle

\section{Introduction}
\label{sec:intro}

Modern machine-learning systems increasingly demand high-performance GPU
kernels. New operators and model architectures, including sparse attention
indexers~\cite{deepseek2025v32exp}, gated delta
networks~\cite{yang2024gateddeltanet}, and latent-attention
variants~\cite{deepseek2024v2}, continue to appear and further expand the
computational patterns that require optimization. Each such pattern is a new
optimization task: work partitioning, dataflow, and pipeline organization
have to be determined again for the concrete operator, input shape, and
hardware.

Faced with this continuing demand for kernel development, agents that
automatically write and optimize operators have drawn attention. Prior work
has built agents dedicated to kernel
optimization~\cite{chen2025nvidiadeepseek, wang2025geak, dong2025stark,
wei2025astra, han2026egg, luo2026ako}, organizing generation, evaluation, and
iterative improvement into a dedicated workflow, and has surpassed expert
implementations on some tasks. These workflows still have two structural
limits. On tasks that fall short of the expert reference, optimization traces
show that architectural decisions, such as work partitioning, dataflow, and
pipeline organization, are often fixed in the early attempts. Subsequent
iterations tune within that architecture and rarely revise these decisions
from performance feedback. In addition, producing
a high-performance kernel in a single run does not mean that its validated
findings have become cross-run knowledge that can guide later architectural
design and implementation.

Overall, an operator-specific agent should close these two limits. First,
performance feedback should revise architectural decisions, rather than
leaving optimization inside an early choice of work partitioning, dataflow,
and pipeline organization. Second, validated findings should become cross-run
knowledge that guides later architectural design and implementation.

\begin{figure}
  \centering
  \includegraphics[width=\columnwidth]{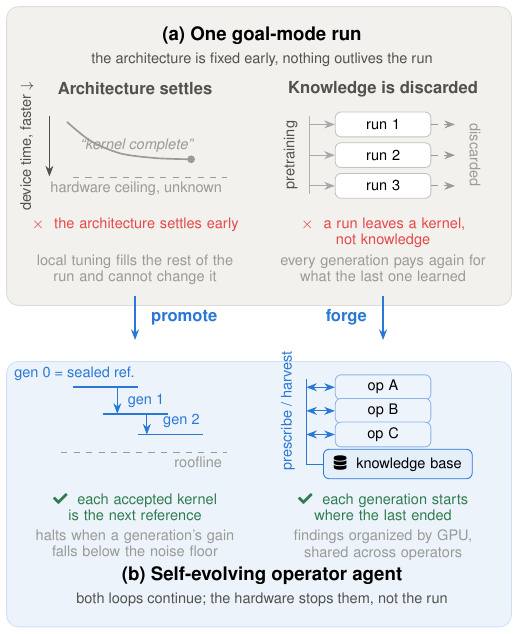}
  \caption{Single-run optimization and cross-run self-evolution.}
  \label{fig:selfevo}
\end{figure}

Guided by these design objectives, we introduce \sys, a self-evolving operator agent
system (Figure~\ref{fig:selfevo}). To address these two problems, \sys starts
optimization from architectural design rather than existing code, and uses
a hardware knowledge base to retain and reuse measurement-validated
optimization experience.

For architectural design, \sys uses a high-performance reference implementation
as a design reference and starting point. It separates architecture from coding,
with a dedicated architecture agent reviewing and adjusting architectural
decisions so that optimization is not confined to code-level tuning within a
fixed architecture. Building on this separation, self-evolution enables
continual reflection on the architecture as a whole and iterative improvement,
reducing the risk of becoming trapped in local optima.

For knowledge accumulation, \sys distills mechanisms discovered during
optimization and validated through independent experiments on the target
hardware into cross-run knowledge that guides architectural design and
implementation in subsequent tasks. Reference updates provide new performance
baselines, while accumulated knowledge provides reusable evidence for
optimization decisions; together, they support continued optimization.

We evaluate \sys on NVIDIA Hopper and Blackwell GPUs across \NumShapes{}
test configurations covering \NumOps{} operators. \sys matches or exceeds
reference performance on \NumBeat{} configurations, achieving a geometric-mean
speedup of \SpeedupRatioMean{}$\times$ over reference implementations from
expert libraries and other kernel-generation systems. On Ascend, \sys
outperforms the per-operator public best on the public case set of all
\AscendOpCount{} level-4 operators in the \cannbench{} Season~\CannSeason{}
final leaderboard. In experiments comparing
against kernels generated by a general-purpose coding agent, \sys achieves
performance gains of up to \GainOverAgent{}. In addition, knowledge-provenance
records demonstrate reuse of validated optimization knowledge across runs and
operators, while knowledge-base ablations show that this knowledge helps
improve generated kernel performance on some evaluated tasks. As external
validation beyond these controlled experiments, as of \LeaderboardSnapshot{},
public submissions generated by \sys rank first on all four leaderboards:
the three \cannbench{}~\CannVTwoVersion{} solution leaderboards
(\CannVTwoBoards{}) and the SOL-ExecBench FlashInfer-Bench track
~\cite{cannbenchleaderboard2026,solexecbenchleaderboard2026}.

The ForgeKernel source code is publicly available at
\url{https://github.com/OpenBMB/ForgeKernel}.

This paper makes the following contributions:

\begin{itemize}
\item We propose a kernel optimization method centered on architectural
decisions. It separates architectural design from code implementation,
guides implementation through explicit architectural decisions, and revises
these decisions based on performance feedback. When optimization encounters
a structural bottleneck, the method revisits work partitioning, dataflow,
or pipeline organization rather than only tuning locally within a fixed
architecture. Across runs, it also reassesses an existing implementation's
architecture rather than directly carrying it forward.

\item We propose a measurement-validated mechanism for accumulating and
reusing knowledge across runs. Hardware mechanisms discovered during
optimization are validated through independent probes on the target GPU
before being stored in a knowledge base shared across operators and input
shapes. This knowledge guides architectural design and implementation
planning in subsequent runs, converting findings from individual runs into
reusable knowledge.

\item We implement \sys, a self-evolving operator optimization system that
integrates architectural revision, reference updates, and knowledge reuse.
Evaluation on NVIDIA Hopper and Blackwell GPUs and Ascend demonstrates
performance gains over expert and generated references, including a
\SpeedupRatioMean{}$\times$ geometric-mean speedup on NVIDIA GPUs.
As of \LeaderboardSnapshot{}, submissions generated by \sys rank first on
all four leaderboards: the three \cannbench{}~\CannVTwoVersion{} solution
leaderboards (\CannVTwoBoards{}) and the SOL-ExecBench FlashInfer-Bench track.
\end{itemize}

\section{Background}
\label{sec:background}

\subsection{Kernel Design}
\label{sec:background:layers}

GPU kernel design requires mapping an operator's computation onto the parallel
execution and memory structures of the target hardware. For a given operator
and input configuration, designers must determine how to partition computation
across thread blocks and threads, lay out data across memory levels, and
pipeline data transfers with computation
~\cite{h100whitepaper,ptxisa,blackwellbrief,cutlass,shah2024flashattention3}.
These choices jointly affect data reuse, memory access overhead, and hardware
resource utilization, so different implementations of the same operator can
exhibit substantial performance differences.

Kernel optimization involves both revising these structural design choices
and refining instructions, synchronization operations, and resource
configurations within a given structure. When performance is limited by work
partitioning, data layout, or pipeline organization, local parameter tuning
alone may be insufficient to eliminate the bottleneck, requiring the
corresponding design choices to be reconsidered. Kernel optimization therefore
involves not only parameter search but also analysis and modification of the
implementation structure.

\subsection{Kernel Agent}
\label{sec:background:agent}

As large language models (LLMs) improve in code generation and reasoning,
researchers have begun using specialized LLM-based agents to write and optimize
kernels. These systems are referred to as kernel agents. Given an operator,
an input configuration, and a target GPU, a kernel agent generates candidate
implementations, compiles them and checks their correctness, measures execution
performance, and iteratively revises the code based on feedback to search for
a functionally correct and high-performance kernel.

\section{Design}
\label{sec:design}

\begin{figure*}
  \centering
  \includegraphics[width=\textwidth]{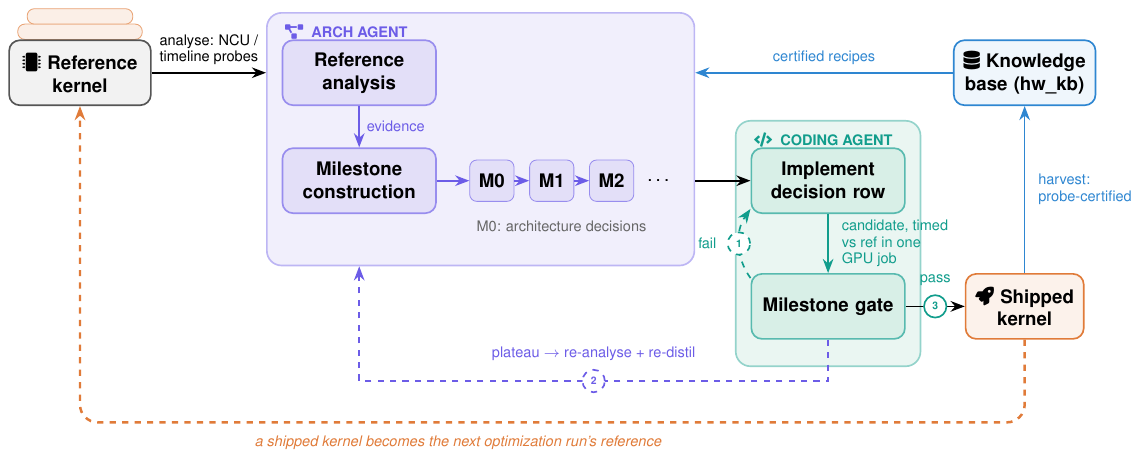}
  \caption{\sys architecture.}
  \label{fig:architecture}
\end{figure*}

\sys separates architectural planning from kernel implementation, coordinating
an architecture agent and a coding agent through design plans and measurement
feedback. Figure~\ref{fig:architecture} shows the system architecture.

\subsection{Overview}
\label{sec:design:overview}
\label{sec:design:contract}

Kernel optimization requires both structural design and detailed implementation.
Structural design determines how computation and data movement are organized
on the target hardware; implementation turns those choices into correct,
executable code. Combining both responsibilities in one agent risks allowing
compilation, debugging, and local edits to dominate the design context and
anchoring further optimization to the current implementation. \sys separates
these responsibilities so that architectural choices can be reconsidered from
measurement evidence without making the existing code structure the starting
point for every revision.

The architecture agent owns design decisions and plan revisions. Given the
operator definition, input configuration, target hardware, and reference
evidence, it determines computation organization, dataflow, scheduling, and
hardware mapping, and specifies what evidence would validate these choices.
Its output is a milestone plan. The coding
agent owns realization of that plan: it implements the required computation,
data access, and synchronization, handles compilation and debugging, and
performs local optimization within the plan's constraints. Changes to those
architectural constraints are handled through plan revision by the architecture
agent.

The interface between the agents is a design plan in one direction and
implementation feedback with measurement evidence in the other. The coding agent receives design
requirements, interfaces, and acceptance conditions, but not the reference
kernel implementation. The execution harness enforces this boundary, checks
correctness, measures performance, and validates the mechanisms required by the
plan. These checks determine whether implementation can advance or needs
further repair; when optimization stalls or the plan cannot proceed, the
trajectory and measurements return to the architecture agent for reassessment.

This separation assigns architectural reasoning, code realization, and measured
acceptance to distinct parts of the system while retaining a feedback loop
between them. Sections~\ref{sec:design:analysis} and~\ref{sec:design:coding}
describe the two agents in detail. Self-evolution extends the loop through plan
revision and reference updates (\S\ref{sec:design:selfref}), while the hardware
knowledge base retains validated mechanisms for reuse
(\S\ref{sec:design:knowledge}).

\subsection{Architecture Agent}
\label{sec:design:analysis}

The architecture agent combines operator requirements, reference measurements, and existing hardware knowledge to formulate architectural decisions and milestone plans. Feedback-driven design revision is described in \S\ref{sec:design:selfref}.

\paragraph{Architectural decisions.}
The architecture agent first determines the candidate implementation's computation organization, dataflow, and mapping to the target hardware. Decisions cover kernel decomposition, data dependencies, and intermediate-result storage, as well as computational tiling, thread roles, register budgets, and shared-memory layouts. Together, these choices specify how computation and data movement interact.

Architectural decisions must state the design choices, supporting evidence, and expected effects, distinguishing direct observations from hypotheses that remain to be tested. The reference implementation's measured performance determines relative performance targets, while its resource usage and execution characteristics inform subsequent mechanism validation. Knowledge entries used in the design are explicitly cited in the plan.

\paragraph{Milestone planning.}
The architecture agent organizes architectural decisions and subsequent optimization tasks into an ordered milestone sequence $M_0, M_1, \ldots, M_k$. The initial milestone $M_0$ requires the implementation of the key structural decisions above to produce a verifiable initial implementation; subsequent milestones perform fine-grained tuning within that structure.

Each milestone specifies implementation tasks, decision rationales, expected costs, and prerequisite dependencies, together with correctness, performance, and mechanism-validation conditions. Dependencies between decisions determine the implementation order, and decisions that must be implemented together to take effect form a single implementation unit. In particular, $M_0$ focuses on correctness and architectural constraints without imposing an absolute performance threshold; subsequent milestones are assessed against their respective optimization objectives. Milestones describe tasks through design requirements and interface contracts, without including the operator's implementation code.

\subsection{Coding Agent}
\label{sec:design:coding}

The coding agent translates design requirements into an operator implementation according to the architecture agent's milestone plan. It follows decision dependencies and consults the knowledge entries cited in the plan to implement the computation, data access, and synchronization required by the current milestone. The harness checks candidate correctness, performs paired timing of the candidate and its declared reference within the same measurement job when that reference can run alongside the candidate, and collects the execution and resource characteristics needed for mechanism validation. Reference-free black-box arms use separate post-stop confirmation channels instead.

The harness determines whether a candidate passes acceptance based on correctness-test results and the milestone's performance and mechanism-validation conditions. For acceptance items that require a performance improvement, the measurements described above determine whether the gain exceeds the measurement noise threshold. Mechanism validation checks the execution or resource characteristics specified in the plan to determine whether the implementation satisfies the corresponding design constraints and exhibits the expected behavioral changes. After acceptance, the coding agent advances to the next milestone; otherwise, it continues modifying the implementation based on the check results. If optimization stagnates or acceptance cannot proceed, the system returns the implementation trajectory and measurement evidence to the architecture agent for the revision process described in \S\ref{sec:design:selfref}.

\subsection{Self-Evolution}
\label{sec:design:selfref}
\label{sec:design:loop}

Self-evolution encompasses plan revision within an optimization run and reference updates across runs.

When optimization stagnates, the executable optimization decisions in the current plan are exhausted without reaching the target, or milestone acceptance cannot proceed, the system triggers reassessment of the current design by the architecture agent. The architecture agent combines implementation trajectories, measurement evidence, and supplementary probe experiments to revise architectural decisions and the corresponding milestone plan.

The revised plan is re-executed according to decision dependencies. Local adjustments resume from the relevant tasks; structural changes to computation organization or dataflow restart implementation from the earliest affected decision, including $M_0$ when necessary. Subsequent implementation and measurement evaluate each revision. If no performance improvement can be attributed to the change, the change is reverted and its failure conditions are recorded to inform subsequent diagnosis.

Beyond revising the current plan, the system can use an accepted implementation as the reference for the next optimization run, extending optimization through its own outputs without requiring a new external expert implementation. This implementation participates in the subsequent run as a measurable black box, with its source code unavailable to that run's agents. The architecture agent analyzes the updated reference and uses it to formulate architectural decisions and performance targets. The coding agent generates a new implementation according to the corresponding milestone plan, while validated knowledge is retained and reused through the knowledge base. Cross-run iteration terminates when the performance gain in a subsequent run is confirmed not to exceed the measurement noise threshold, or when the budget is exhausted.

\subsection{Knowledge Validation and Reuse}
\label{sec:design:knowledge}

The hardware knowledge base stores hardware mechanisms distilled during optimization and diagnosis, organized by target hardware architecture. Each entry records the mechanism, applicability conditions, trigger symptoms, a minimal runnable example, and known limitations. A candidate mechanism is admitted after passing independent probe validation on the target hardware and providing supporting measurement evidence.

During subsequent optimization, the architecture agent selects relevant knowledge based on the task's measured characteristics and the entries' applicability conditions, and explicitly cites the entries in milestones for the coding agent to implement the corresponding mechanisms. Milestone acceptance still checks their effects in the current task. The system records the creation and use of entries to support traceability and reuse across runs.

\section{Evaluation}
\label{sec:eval}

This chapter evaluates the performance of \sys-generated implementations
and the roles of architecture planning, feedback-driven revision, and knowledge
reuse. \S\ref{sec:eval:setup} describes the setup; \S\ref{sec:eval:gap} reports
public-benchmark results and comparisons with expert libraries and artifacts
from other generation systems; \S\ref{sec:eval:selfevo} evaluates self-evolution;
\S\ref{sec:eval:ablation} examines the individual designs; and
\S\ref{sec:eval:case-study} uses GQA and DSA cases to examine the roles of the
architecture agent and self-evolution in architecture changes.

\subsection{Experimental Setup}
\label{sec:eval:setup}

\emph{Hardware and tasks.} NVIDIA experiments run on Hopper H100 80\,GB
(SXM, 132 SMs) and Blackwell B200 (148 SMs) nodes in a shared cluster,
covering attention, matrix multiplication, convolution, GDN, and DSA.
Ascend experiments run on \AscendHw{}; the per-operator comparison covers
all \AscendOpCount{} level-4 operators in
\cannbench{}~\AscendBenchVersion{}. Coverage is reported separately for
public leaderboards and per-operator experiments. Input shapes and data
types are listed in Appendix~\ref{sec:app:shapes}.

\emph{Comparison targets.} Per-operator experiments use expert library
implementations and kernels released by other generation systems as
references. Libraries include FlashAttention-4, DeepGEMM, cuBLAS, cuDNN,
FlashInfer, FlashMLA, and FlashQLA; generation systems include K-Search,
Kernel Design Agents, and CAKE. Reference versions are pinned in the execution
environment image and re-timed in our environment. Comparisons with generation
systems evaluate their released artifacts.
Versions and entry points are listed in Appendix~\ref{sec:app:refs}.

\emph{Models and run protocol.} The arch agent and coding agent of \sys
(\S\ref{sec:design:overview}) use Claude Fable~5.1 and Claude Opus~5,
respectively. By default, \sys receives a sealed reference that can be run,
timed, profiled, and disassembled, without source access.
% Exceptions in the reference-information ablation are specified in \S\ref{sec:eval:ablation}.
The knowledge base accumulates across runs, so its initial state can differ
between runs. Performance targets and budgets are set per task; runs end
when the target is reached, the budget is exhausted, or the run is stopped
manually. Expert-library experiments target parity, while self-evolution
typically targets \SelfEvoTarget{} over the preceding generation; specific
targets and stopping reasons follow the individual run records. Main
performance results select the best optimization run per configuration.
The full system and each variant in the module ablation are run separately
for that experiment.

\emph{Correctness validation.} Candidates must pass the official or benchmark
tolerance test associated with their operator source before performance
measurement. Operators covered by FlashInfer-Bench, including the operators
in the SOL-ExecBench FlashInfer-Bench collection, use FlashInfer-Bench's
tolerance tests; FlashAttention operators use the official FlashAttention
tolerance tests; other operators likewise use the tolerance criterion of
their source benchmark or implementation. The harness also rejects candidates
that call the reference
implementation directly or delegate computation to it. The source-specific
criteria are summarized in Appendix~\ref{sec:app:judges}.

\emph{Timing method.} On NVIDIA hardware, we measure device execution time
under cold-\(L2\) cache conditions. Average device time per call is the total
GPU time over repeated calls divided by the call count. The candidate and
reference use the same timing protocol and cache conditions. Results are
based on repeated measurements. Speedup is
\(S=T_{\mathrm{ref}}/T_{\mathrm{candidate}}\), where the numerator and denominator
are the device execution times of the reference and candidate, respectively;
performance gain is \((S-1)\times100\%\).

Ascend experiments use the hosted evaluation harness of \cannbench{}, with
device execution time recorded in msprof's \texttt{kernel\_details}.
Baselines, hardware performance bounds, and scoring rules follow the
corresponding benchmark version. Scores are computed by the hosted service;
results are aggregated separately for each platform.

\emph{The \texttt{/goal} configuration.} The \goalagent{} follows the
goal-driven setting of JustGoal~\cite{chen2026justgoal}, using Claude
Fable~5.1 without an operator-specific workflow and with a 24-hour budget
per task. Prompts use a fixed format specifying the operator signature,
input shapes, target hardware (one H100), and optimization goal; copying
code from the web is prohibited. No existing implementation is supplied as a
reference. Generated implementations are subsequently evaluated against expert
libraries. The objective has no preset speedup ceiling.
Optimization state is not retained across sessions, and acquired knowledge
is not transferred between runs. Selected artifacts serve as initial
references for some self-evolution experiments in \S\ref{sec:eval:selfevo}.

\subsection{Performance Comparison}
\label{sec:eval:gap}

\paragraph{Overall performance on public benchmarks.}
As of \LeaderboardSnapshot{}, public submissions generated by \sys rank first
on all three \cannbench{}~\CannVTwoVersion{} solution leaderboards---
\CannVTwoBoards{}---and on the SOL-ExecBench FlashInfer-Bench collection
leaderboard~\cite{cannbenchleaderboard2026,solexecbenchleaderboard2026}.
Table~\ref{tab:public-benchmarks} reports coverage, speedups, and ranks.
Speedups use each benchmark's own baseline and aggregation rule.
Ranks follow the official scores reported below.

For the detailed historical result retained in this paper, on the
\cannbench{} Season~\CannSeason{} A3 (Ascend \CannLeaderboardHw{})
\CannLeaderboardVersion{} leaderboard, \sys covers all
\CannLeaderboardOps{} operators across levels 1--4, including elementwise
operations, normalization, matrix multiplication, convolution, attention,
and fused operators. It passes all \CannLeaderboardCases{} standard-set
cases and \CannHiddenPassed{} of \CannHiddenCases{} hidden-set cases.
The two sets score \CannStandardScore{} and \CannHiddenScore{}, respectively,
yielding a first-place total of \CannLeaderboardScore{}. Geometric-mean
speedups over the benchmark reference are \(\CannStandardSpeedup\times\)
on the standard set and \(\CannHiddenSpeedup\times\) on the hidden set;
the leaderboard reports a combined baseline speedup of
\(\CannCombinedSpeedup\times\), computed as their weighted geometric mean
with weights of \(20\%\) and \(80\%\), respectively
~\cite{cannbenchleaderboard2026}.

On SOL-ExecBench, the FlashInfer-Bench collection covers normalization,
matrix multiplication, GQA/MLA attention, and MoE kernels. With evaluation
stack \SolExecStack{} on NVIDIA \SolExecHw{}, \sys solves all
\SolExecProblems{} kernels, spanning \SolExecWorkloads{} workloads, and
achieves a SOL Score of \SolExecScore{}. Beyond this first-place score,
\(\mathrm{Fast}_1=\SolExecFastCount/\SolExecWorkloads\), or
\SolExecFastPercent{}, shows that the generated implementations outperform
the scoring baseline on almost all workloads. The leaderboard reports an
\emph{Avg Speedup} of \(\SolExecAvgSpeedup\times\) over that baseline,
providing a complementary measure of acceleration magnitude rather than
only the fraction of workloads improved~\cite{solexecbenchleaderboard2026}.

Here, \(\mathrm{Fast}_1\) is the fraction of workloads strictly faster than
the baseline, aggregated across kernels.

\begin{table}[t]
  \centering
  \fontsize{8.5}{10.5}\selectfont
  \caption{Public-benchmark results.}
  \label{tab:public-benchmarks}
  \setlength{\tabcolsep}{2pt}
  \renewcommand{\arraystretch}{1.1}
  \begin{tabular*}{\columnwidth}{@{\extracolsep{\fill}}
    >{\raggedright\arraybackslash}p{0.27\columnwidth}
    >{\raggedright\arraybackslash}p{0.42\columnwidth}
    >{\raggedleft\arraybackslash}p{0.16\columnwidth}
    >{\raggedleft\arraybackslash}p{0.09\columnwidth}@{}}
  \toprule
  \textbf{Benchmark} & \textbf{Coverage} & \textbf{Speedup} & \textbf{Rank} \\
  \midrule
  \textbf{\cannbench{} v\CannLeaderboardVersion{}}\newline Ascend \CannLeaderboardHw{}
    & \CannLeaderboardOps{} operators\newline Standard: \CannLeaderboardCases{}/\CannLeaderboardCases{}\newline Hidden: \CannHiddenPassed{}/\CannHiddenCases{}
    & \(\pgfmathprintnumber[fixed,precision=2,fixed zerofill]{\CannCombinedSpeedup}\times\) & \textbf{1} \\
  \addlinespace[4pt]
  \textbf{SOL-ExecBench}\newline NVIDIA \SolExecHw{}
    & \SolExecProblems{}/\SolExecProblems{} kernels\newline \SolExecWorkloads{} workloads
    & \(\pgfmathprintnumber[fixed,precision=2,fixed zerofill]{\SolExecAvgSpeedup}\times\) & \textbf{1} \\
  \bottomrule
  \end{tabular*}
\end{table}

\paragraph{Operator-level performance comparison.}
Beyond the overall public-benchmark results, we evaluate representative
operators to compare \sys with expert-library implementations, artifacts
from existing generation systems, and per-operator public-best submissions.

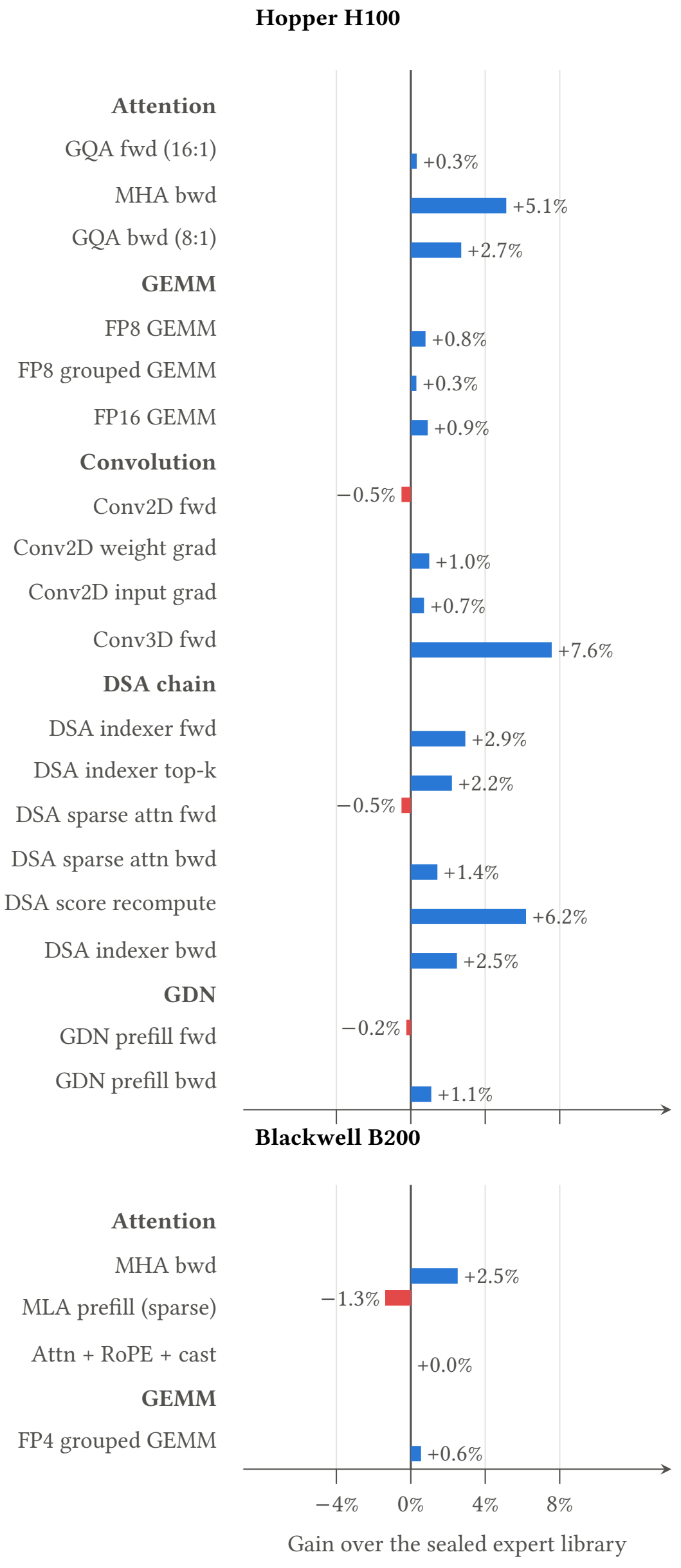
\begin{figure}[t]
  \centering
  % AUTO-GENERATED by scripts/gen_figures.py -- edit data/*.csv instead.
\begin{tikzpicture}
\begin{groupplot}[
  fkgainbar, width=0.52\columnwidth, y=13pt,
  group style={group size=1 by 2, vertical sep=22pt,
               xlabels at=edge bottom, xticklabels at=edge bottom},
  xmin=-9, xmax=14, xtick={-4,0,4,8},
  xlabel={Gain over the sealed expert library},
]
\nextgroupplot[
  title={Hopper H100},
  title style={font=\scriptsize\bfseries, anchor=south west, at={(0,1)}, yshift=2pt},
  ymin=-0.8, ymax=22.6,
  ytick={0,1,2,3,4,5,6,7,8,9,10,11,12,13,14,15,16,17,18,19,20,21,22},
  yticklabels={{\textbf{Attention}},{GQA fwd (16:1)},{MHA bwd},{GQA bwd (8:1)},{\textbf{GEMM}},{FP8 GEMM},{FP8 grouped GEMM},{FP16 GEMM},{\textbf{Convolution}},{Conv2D fwd},{Conv2D weight grad},{Conv2D input grad},{Conv3D fwd},{\textbf{DSA chain}},{DSA indexer fwd},{DSA indexer top-k},{DSA sparse attn fwd},{DSA sparse attn bwd},{DSA score recompute},{DSA indexer bwd},{\textbf{GDN}},{GDN prefill fwd},{GDN prefill bwd}},
]
\draw[fkInk, semithick] (axis cs:0,-0.8) -- (axis cs:0,22.6);
\addplot[draw=none, fill=fkGain, nodes near coords, point meta=explicit symbolic,
  every node near coord/.append style={anchor=west}]
  coordinates {(0.32,1) [+0.3\%] (5.13,2) [+5.1\%] (2.71,3) [+2.7\%] (0.79,5) [+0.8\%] (0.3,6) [+0.3\%] (0.91,7) [+0.9\%] (0.99,10) [+1.0\%] (0.71,11) [+0.7\%] (7.57,12) [+7.6\%] (2.93,14) [+2.9\%] (2.21,15) [+2.2\%] (1.43,17) [+1.4\%] (6.19,18) [+6.2\%] (2.48,19) [+2.5\%] (1.1,22) [+1.1\%]};
\addplot[draw=none, fill=fkLoss, nodes near coords, point meta=explicit symbolic,
  every node near coord/.append style={anchor=east}]
  coordinates {(-0.5,9) [$-$0.5\%] (-0.5,16) [$-$0.5\%] (-0.24,21) [$-$0.2\%]};
\nextgroupplot[
  title={Blackwell B200},
  title style={font=\scriptsize\bfseries, anchor=south west, at={(0,1)}, yshift=2pt},
  ymin=-0.8, ymax=5.6,
  ytick={0,1,2,3,4,5},
  yticklabels={{\textbf{Attention}},{MHA bwd},{MLA prefill (sparse)},{Attn + RoPE + cast},{\textbf{GEMM}},{FP4 grouped GEMM}},
]
\draw[fkInk, semithick] (axis cs:0,-0.8) -- (axis cs:0,5.6);
\addplot[draw=none, fill=fkGain, nodes near coords, point meta=explicit symbolic,
  every node near coord/.append style={anchor=west}]
  coordinates {(2.52,1) [+2.5\%] (0,3) [+0.0\%] (0.55,5) [+0.6\%]};
\addplot[draw=none, fill=fkLoss, nodes near coords, point meta=explicit symbolic,
  every node near coord/.append style={anchor=east}]
  coordinates {(-1.35,2) [$-$1.3\%]};
\end{groupplot}
\end{tikzpicture}
  \caption{Performance gains over expert libraries on Hopper and Blackwell.}
  \label{fig:forge-results}
\end{figure}

\begin{figure}[t]
  \centering
  % AUTO-GENERATED by scripts/gen_figures.py -- edit data/*.csv instead.
\begin{tikzpicture}
\begin{axis}[
  fkgainbar, width=0.44\columnwidth, y=13pt,
  ymin=-0.8, ymax=8.6,
  xmin=-14, xmax=200, xtick={0,50,100,150},
  xlabel={Gain over the sealed system kernel},
  ytick={0,1,2,3,4,5,6,7,8},
  yticklabels={{\makebox[0.32\columnwidth][l]{\textbf{Hopper H100}}\hspace{4pt}\makebox[0.12\columnwidth][l]{\fontsize{6.5}{8}\selectfont\textcolor{black!55}{\textbf{Ref.}}}},{\makebox[0.32\columnwidth][l]{GQA decode (paged)}\hspace{4pt}\makebox[0.12\columnwidth][l]{\fontsize{6.5}{8}\selectfont\textcolor{black!55}{K-Search}}},{\makebox[0.32\columnwidth][l]{MLA decode (paged)}\hspace{4pt}\makebox[0.12\columnwidth][l]{\fontsize{6.5}{8}\selectfont\textcolor{black!55}{K-Search}}},{\makebox[0.32\columnwidth][l]{MLA prefill (paged)}\hspace{4pt}\makebox[0.12\columnwidth][l]{\fontsize{6.5}{8}\selectfont\textcolor{black!55}{K-Search}}},{\makebox[0.32\columnwidth][l]{TriMul}\hspace{4pt}\makebox[0.12\columnwidth][l]{\fontsize{6.5}{8}\selectfont\textcolor{black!55}{K-Search}}},{\makebox[0.32\columnwidth][l]{\textbf{Blackwell B200}}\hspace{4pt}\makebox[0.12\columnwidth][l]{\fontsize{6.5}{8}\selectfont\textcolor{black!55}{\textbf{Ref.}}}},{\makebox[0.32\columnwidth][l]{DSA sparse attn (paged)}\hspace{4pt}\makebox[0.12\columnwidth][l]{\fontsize{6.5}{8}\selectfont\textcolor{black!55}{KDA}}},{\makebox[0.32\columnwidth][l]{GDN prefill (varlen)}\hspace{4pt}\makebox[0.12\columnwidth][l]{\fontsize{6.5}{8}\selectfont\textcolor{black!55}{KDA}}},{\makebox[0.32\columnwidth][l]{KDA prefill (recurrent)}\hspace{4pt}\makebox[0.12\columnwidth][l]{\fontsize{6.5}{8}\selectfont\textcolor{black!55}{CAKE}}}},
]
\draw[fkInk, semithick] (axis cs:0,-0.8) -- (axis cs:0,8.6);
\addplot[draw=none, fill=fkGain, nodes near coords, point meta=explicit symbolic,
  every node near coord/.append style={anchor=west}]
  coordinates {(24.47,1) [+24.5\%] (170.6,2) [+170.6\%] (146.1,3) [+146.1\%] (8.71,4) [+8.7\%] (7.03,6) [+7.0\%] (5.82,7) [+5.8\%] (0.37,8) [+0.4\%]};
\addplot[draw=none, fill=fkLoss, nodes near coords, point meta=explicit symbolic,
  every node near coord/.append style={anchor=east}]
  coordinates {};
\end{axis}
\end{tikzpicture}
  \caption{Performance gains over published kernels from other generation systems.}
  \label{fig:forge-systems}
\end{figure}
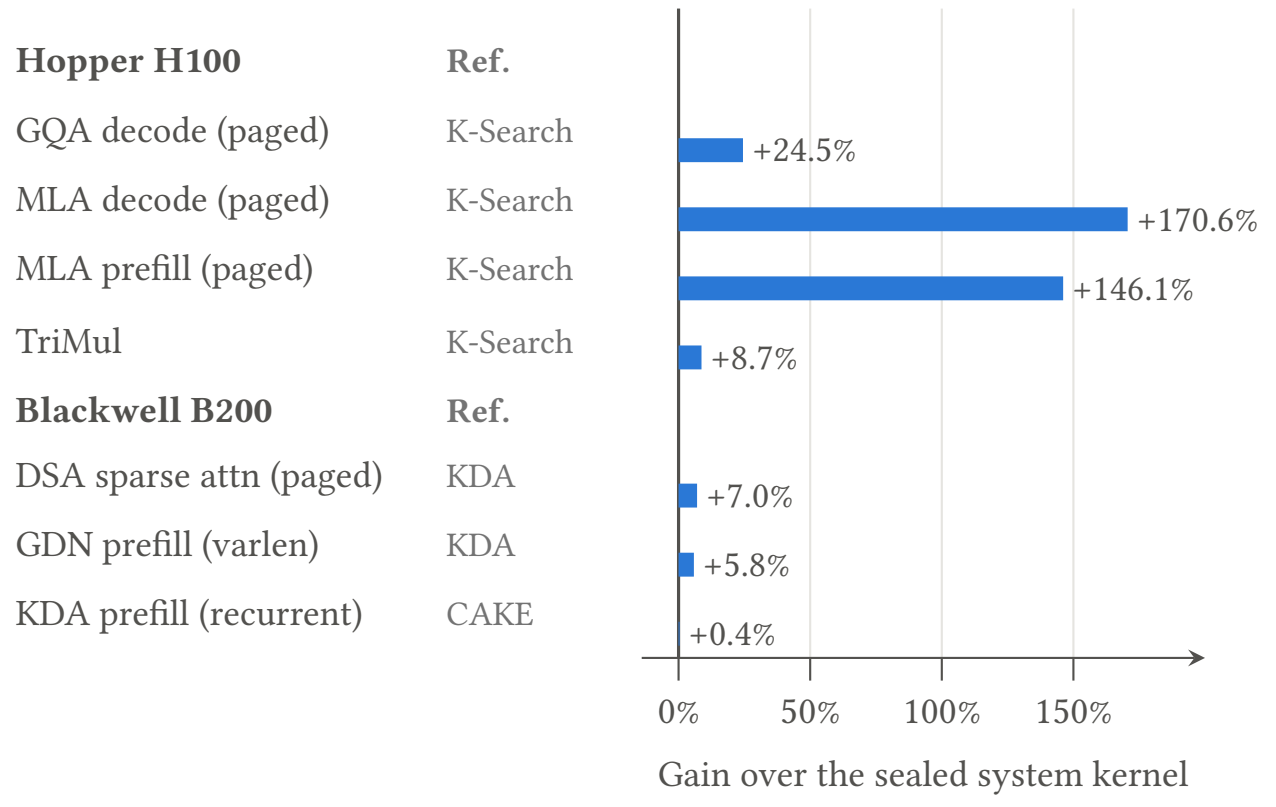

Across \NumLibraryCells{} configurations on Hopper and Blackwell, \sys
matches or exceeds the expert implementation in \NumLibraryBeat{}, with
another \NumLibraryNear{} within 1\% of the reference. The geometric-mean
performance gain across all configurations is \SpeedupMeanLibrary{}
(Figure~\ref{fig:forge-results}).
Each bar in Figures~\ref{fig:forge-results} and~\ref{fig:forge-systems}
represents the best optimization run for an operator, input shape, and hardware
configuration, using the measurement protocol in \S\ref{sec:eval:setup}.
The vertical zero line marks parity; the Hopper and Blackwell panels in
Figure~\ref{fig:forge-results} share a scale.

Beyond expert libraries, \sys matches or exceeds all \NumSystemCells{} published kernels from
K-Search~\cite{cao2026ksearch}, Kernel Design Agents~\cite{liu2026kda},
and CAKE~\cite{ye2026cake}, with a geometric-mean gain of
\SpeedupMeanSystem{} (Figure~\ref{fig:forge-systems}).
Hopper references are K-Search's FlashInfer-Bench and TriMul implementations;
Blackwell references are KDA-v0.5~\cite{kdacontest2026} and CAKE implementations.

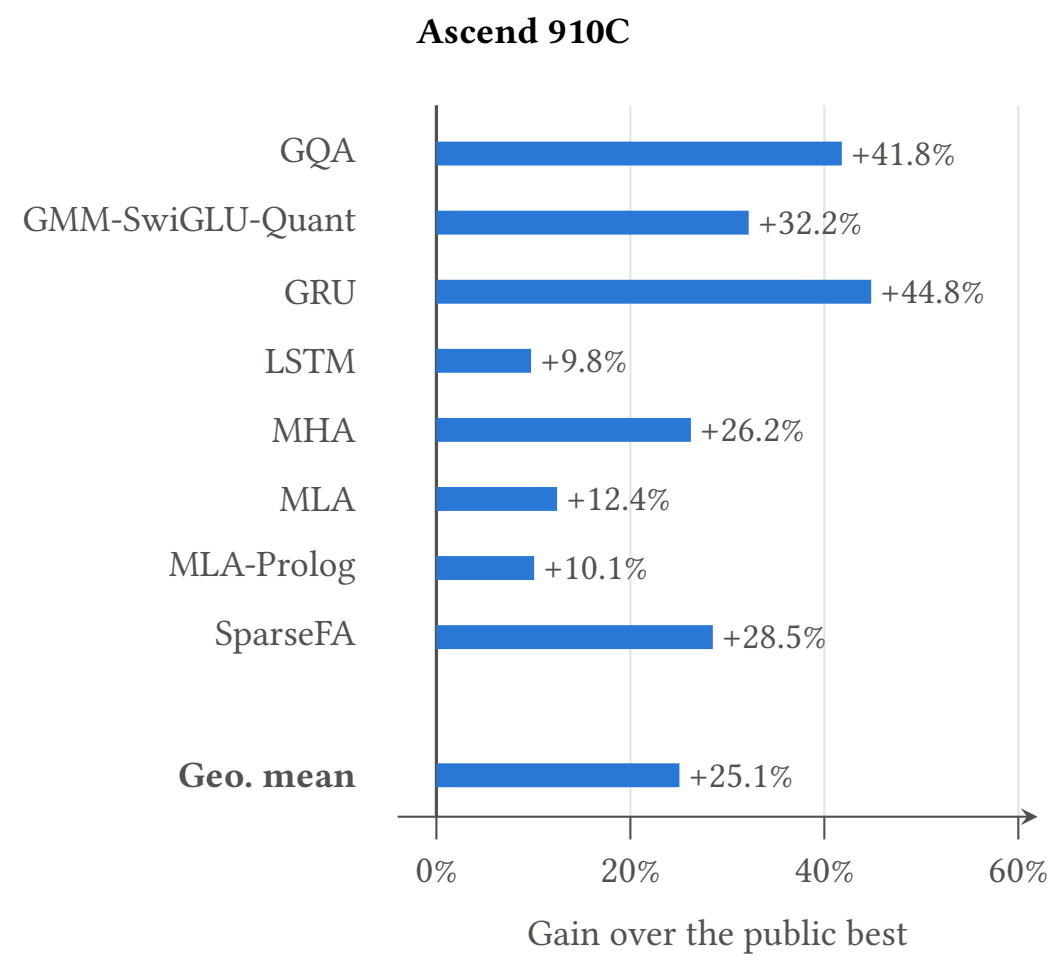
\begin{figure}[t]
  \centering
  % AUTO-GENERATED by scripts/gen_figures.py -- edit data/*.csv instead.
\begin{tikzpicture}
\begin{axis}[
  fkgainbar, width=0.50\columnwidth, y=13pt,
  title={Ascend \AscendHw{}},
  title style={font=\scriptsize\bfseries, anchor=south west, at={(0,1)}, yshift=2pt},
  ymin=-0.7, ymax=9.6,
  xmin=-4, xmax=62, xtick={0,20,40,60},
  xlabel={Gain over the public best},
  ytick={0,1,2,3,4,5,6,7,9},
  yticklabels={{GQA},{GMM-SwiGLU-Quant},{GRU},{LSTM},{MHA},{MLA},{MLA-Prolog},{SparseFA},{\textbf{Geo. mean}}},
]
\draw[fkInk, semithick] (axis cs:0,-0.7) -- (axis cs:0,9.6);
\addplot[draw=none, fill=fkGain, nodes near coords, point meta=explicit symbolic,
  every node near coord/.append style={anchor=west}]
  coordinates {(41.8,0) [+41.8\%] (32.2,1) [+32.2\%] (44.83,2) [+44.8\%] (9.756,3) [+9.8\%] (26.24,4) [+26.2\%] (12.44,5) [+12.4\%] (10.08,6) [+10.1\%] (28.5,7) [+28.5\%] (25.06,9) [+25.1\%]};
\end{axis}
\end{tikzpicture}
  \caption{Performance gains over public-best submissions on Ascend \AscendHw{}.}
  \label{fig:ascend}
\end{figure}

We further evaluate \sys on Ascend \AscendHw{}, using the public case sets of all
\AscendOpCount{} level-4 operators in the \cannbench{} Season~\CannSeason{}
final leaderboard. The \sys-generated implementations outperform the
per-operator public best, with a geometric-mean performance gain of
\AscendGeomeanLead{} (Figure~\ref{fig:ascend}).
Each bar reports the percentage performance gain over the per-operator public
best, computed from the ratio of their speedups against the same baseline.
The zero line marks parity; the final row reports the geometric-mean speedup
converted to a percentage gain. Public-best submissions come from the same
final leaderboard and are not optimization inputs. The benchmark evaluates
20 public cases per operator, weighting compilation, accuracy, and
hardware-attainable performance by \(0.2\), \(0.3\), and \(0.5\), respectively.

\subsection{Self-Evolution Experiments}
\label{sec:eval:selfevo}

This section evaluates reference updates across runs: an existing implementation
becomes the next run's reference. In the default \sys protocol, this reference
is measurable but source-hidden, and the arch agent analyzes it to develop a
new plan for the coding agent to implement. The DeepSeek-V4.1 case below uses
a source-visible generated reference and a single coding agent, as specified
in its case description.
We evaluate two paths. In the first, an initial implementation is generated
without any existing implementation as a reference; \sys then
uses that artifact as its reference for further optimization. In the second,
\sys first uses an expert-library implementation as its reference to generate
generation~0, then uses the preceding generation as the reference for
self-evolution. Generation~0 is the output of the first optimization run.

Table~\ref{tab:selfevo} reports initial performance, subsequent round gains,
and final cumulative gains. Initial performance is expressed as a percentage
gain relative to the expert library; final gains use the same expert
baseline. Round~1 is relative to the initial implementation, and round~2 to the
round~1 output. Final gains for the DeepSeek-V4.1 case are measured directly
against FlashMLA; the remaining final cumulative gains are estimated by
compounding the initial and subsequent speedups. A dash denotes a round that
was not conducted.

For the path beginning with reference-free generation, starting from the
\texttt{/goal} implementations of GDN prefill
forward and GQA attention forward, \sys obtains further gains of
\SelfEvoGdnFwdAgentrefGenOne{} and \SelfEvoAttnFwdAgentrefGenOne{}, respectively.
Combined with the initial implementations' performance relative to expert
libraries, the cumulative gains are approximately \SelfEvoCumulGdnFwdAgentref{}
and \SelfEvoCumulAttnFwdAgentref{}.

For the path beginning with an expert-library reference, the initial optimization
is followed by one or two self-evolution rounds. GQA attention forward involves
three \sys runs in total: the initial run produces generation~0, and the next
two produce generations~1 and~2, with cumulative gains over the expert library
of approximately +10\% and \SelfEvoCumulAttnFwdGqa{}, respectively.
This sequence is independent of the GQA experiment beginning with reference-free
generation.

\begin{table*}[t]
  \centering
  \small
  \caption{Initial performance and subsequent gains along two optimization paths.}
  \label{tab:selfevo}
  \renewcommand{\arraystretch}{1.15}
  \setlength{\tabcolsep}{10pt}
  \begin{tabular}{@{}lr>{\bfseries}r>{\bfseries}rr@{}}
  \toprule
  Operator & \shortstack[r]{Initial performance\\vs. expert} & \shortstack[r]{Round 1 gain\\vs. initial} & \shortstack[r]{Round 2 gain\\vs. round 1} & \shortstack[r]{Final gain\\vs. expert} \\
  \midrule
  \multicolumn{5}{@{}l}{\textbf{Reference-free generation, then optimization} (\sys optimizes the generated implementation)} \\
  \addlinespace[2pt]
  GDN prefill fwd & \SelfEvoBaseGdnFwdAgentref{} & \SelfEvoGdnFwdAgentrefGenOne{} & \textnormal{---} & \SelfEvoCumulGdnFwdAgentref{} \\
  GQA fwd (16:1) & \SelfEvoBaseAttnFwdAgentref{} & \SelfEvoAttnFwdAgentrefGenOne{} & \textnormal{---} & \SelfEvoCumulAttnFwdAgentref{} \\
  DeepSeek-V4.1 fused sparse-attention prefill & \SelfEvoDsvFusedInitial{} & \SelfEvoDsvFusedGenOne{} & \textnormal{---} & \SelfEvoDsvFusedCumul{} \\
  \midrule
  \multicolumn{5}{@{}l}{\textbf{Expert-reference optimization, then self-evolution} (\sys throughout)} \\
  \addlinespace[2pt]
  DSA sparse attn fwd & \SelfEvoBaseDsaSparseFwd{} & \SelfEvoDsaSparseFwdGenOne{} & \textnormal{---} & \SelfEvoCumulDsaSparseFwd{} \\
  GDN prefill bwd & \SelfEvoBaseGdnPrefillBwd{} & \SelfEvoGdnPrefillBwdGenOne{} & \textnormal{---} & \SelfEvoCumulGdnPrefillBwd{} \\
  GQA fwd (16:1) & \SelfEvoAttnFwdGqaGenZero{} & \SelfEvoAttnFwdGqaGenOne{} & \SelfEvoAttnFwdGqaGenTwo{} & \SelfEvoCumulAttnFwdGqa{} \\
  \bottomrule
  \end{tabular}
\end{table*}

\paragraph{Self-evolution on a complex fused operator.}
We examine self-evolution on a complex fused operator using DeepSeek-V4.1
fused sparse-attention prefill. Its fused structure provides a demanding test
of \sys's capability.
The experiment uses a single coding agent (Claude Opus~5.5, high effort)
for approximately 24 hours per generation
on B200. The workload fixes \(s_q=4096\), \(h_q=128\),
\(s_{\mathrm{kv}}=32768\), and \(\mathrm{topk}=640\), with an attention sink.
FlashMLA is used for evaluation only; the agent cannot inspect its source or
run it during generation and optimization. The first generation is produced
in \texttt{/goal} mode. The next generation can read the first
generation's source, but not its optimization notes.
The initial gain of \SelfEvoDsvFusedInitial{} over FlashMLA is the median of
three independent same-card retests. The first-round gain
\SelfEvoDsvFusedGenOne{} and final gain \SelfEvoDsvFusedCumul{} are the separate
medians of four acceptance measurements of the final implementation against
the initial implementation and FlashMLA, respectively.
All four final acceptances pass the five official correctness cases and
the in-place-overwrite re-check. The task-specific benchmark reports
cold-\(L2\) device time.
The operator combines Q RoPE, sparse MQA attention with an attention sink,
conjugate output RoPE, and FP8 casting. The CUDA C++ implementation
fuses these operations into one kernel on B200. Its architecture uses a
persistent pair of CTAs to process each query token, cooperatively gathers KV
data, and assigns separate warp roles to data movement, matrix multiplication,
softmax, and the epilogue.

Specific architecture changes during cross-run optimization and their decision
processes are examined in \S\ref{sec:eval:case-study}.

\subsection{Ablation Study}
\label{sec:eval:ablation}

This section examines architecture planning and revision,
disassembly, and the knowledge base. We evaluate within-run plan revision
separately from cross-run reference updates, whose results appear in
\S\ref{sec:eval:selfevo}. The experimental conditions and results follow.

\paragraph{Architecture planning and revision.}
Table~\ref{tab:module-ablation} compares the full system with configurations
that individually remove the milestone template, plan revision, SASS disassembly,
or the knowledge base. Each cell reports the performance gain over the expert
reference from one optimization run. Retaining only the initial
plan without revision reduces performance on all three tasks relative to the
full system, supporting the role of feedback-driven revision on these tasks.
Removing the milestone template substantially reduces GDN prefill-forward
performance, with a smaller change for DSA sparse attention forward. The
attention-backward implementation without the template passes output-error
checks but lowers internal precision (marked $^\dagger$). The DSA run without
the knowledge base uses different inputs, and its result is reported separately.

\begin{table}
  \centering
  \small
  \caption{Performance gains over expert references in component ablations.}
  \label{tab:module-ablation}
  \resizebox{\columnwidth}{!}{% AUTO-GENERATED by scripts/gen_forge_tables.py -- edit data/*.csv instead.
\begin{tabular}{@{}lrrrr@{}}
\toprule
Configuration & \shortstack{Attn\\bwd} & \shortstack{GDN\\prefill fwd} & \shortstack{DSA sparse\\fwd} & \shortstack{Geo.\\mean} \\
\midrule
Full \sys & +0.6\% & $-$1.5\% & +2.0\% & +0.4\% \\
No milestone template & +16.2\%$^\dagger$ & $-$16.2\% & +1.4\% & $-$0.4\% \\
First milestone only & $-$1.0\% & $-$4.7\% & $-$1.2\% & $-$2.3\% \\
No SASS (NCU kept) & $-$10.2\% & +16.2\% & $-$5.1\% & $-$0.3\% \\
No knowledge base & +0.3\% & $-$16.4\% & +6.2\% & $-$3.8\% \\
\bottomrule
\end{tabular}
}
\end{table}

\paragraph{Disassembly information.}
Removing SASS while retaining NCU profiling reduces performance on attention
backward and DSA sparse attention forward, whereas GDN prefill forward exceeds
the full system.

\begin{figure*}
  \centering
  \includegraphics[width=0.9\textwidth]{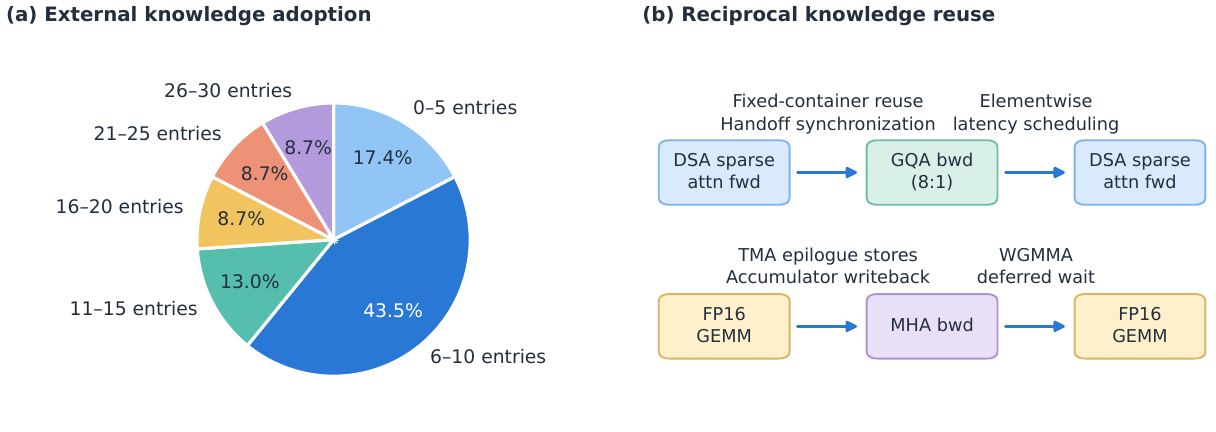}
  \caption{External-knowledge adoption and reciprocal reuse.}
  \label{fig:kb-provenance}
\end{figure*}

\paragraph{Knowledge base.}
\label{sec:eval:provenance}
Knowledge reuse is widespread
(Figure~\ref{fig:kb-provenance}a). Nearly all configurations in the SM90
provenance cohort adopt external knowledge. The most common adoption range is
6--10 distinct entries, while the highest reaches 26--30: operators typically
draw on multiple entries, and the extent of reuse varies substantially,
consistent with the differing ablation effects.

Reuse also crosses operator boundaries in both directions
(Figure~\ref{fig:kb-provenance}b). In the DSA--GQA and GEMM--MHA cases, new
mechanisms developed by a knowledge recipient are adopted in subsequent runs
of the original contributor, establishing cross-operator knowledge feedback.
The knowledge base therefore serves as a shared, evolving collection
of hardware mechanisms. Counting rules and the detailed source--recipient matrix
are provided in Appendix~\ref{sec:app:kb-provenance}.

\subsection{Case Studies: Two Mechanisms for Architecture Changes}
\label{sec:eval:case-study}

The two cases illustrate the roles of the architecture agent and self-evolution
in architecture changes. By separating architecture from coding, the
architecture agent explicitly plans and revises interdependent architectural
decisions. The self-evolution case focuses on using an accepted implementation
as the reference for a subsequent run, with renewed analysis and design to
explore further architectural improvements. These roles are complementary:
the former organizes architecture changes, while the latter makes an existing
optimized artifact the starting point for further architecture exploration.

\paragraph{Architecture-agent-driven replanning: GQA.}
The first GQA attention-forward self-evolution round illustrates how the
architecture agent coordinates interdependent changes. The initial
implementation assigns one query block to each thread block (CTA), which
then exits. Rather than proposing only local code changes, the architecture
agent changes this to persistent execution, so that the same CTA processes
multiple query blocks. It organizes cross-block state management and Q-data
prefetching into a staged plan for the coding agent to implement.

The coding agent first implements the persistent loop and cross-block state
management, obtaining a local performance gain of approximately
\GqaCasePersistentGain{}. It then adds Q double buffering to load the next
query block's data early, obtaining a further local gain of approximately
\GqaCaseQSlotsGain{}. Both steps undergo correctness checks and separate
performance measurements; their local gains cannot be directly added.

The round ultimately improves performance over the preceding generation by
\SelfEvoAttnFwdGqaGenOne{} (Table~\ref{tab:selfevo}), including other changes;
this gain cannot be attributed entirely to the two steps above. The case
highlights how separating architecture from coding supports coordinated
planning: the initial structural change delivers only a modest gain, but
enables subsequent pipeline optimization, with the coding agent implementing
the interdependent changes in dependency order.

\paragraph{Architecture reassessment through self-evolution: DSA.}
The DSA sparse attention-forward case examines how an existing optimized
artifact becomes the starting point for a new round of architecture analysis.
The initial implementation uses non-persistent execution with one resident
CTA per SM. After this implementation becomes the reference, the next round
obtains approximately 4.5\% gain within the original architecture, after which
further adjustments provide no effective improvement. Updating the reference
does not require retaining the preceding generation's architecture, allowing
the system to reassess its computation organization rather than only tune
parameters within that structure.

\begin{figure}[t]
  \centering
  \includegraphics[width=\columnwidth]{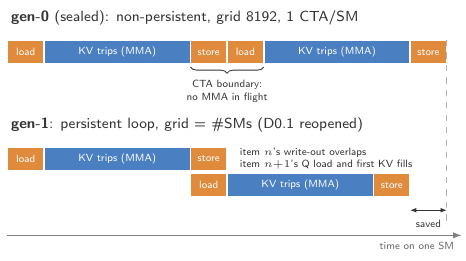}
  \caption{DSA sparse attention-forward architecture change (schematic, not to scale).}
  \label{fig:selfevo-dsa}
\end{figure}

Through diagnostic experiments, the architecture agent finds that each CTA's
query load, initial KV fills, and final output store lack computation from
adjacent work items to overlap with. A probe removing output stores changes
performance by approximately 6.0\%, helping isolate uncovered overhead; this
probe is diagnostic, not a complete candidate implementation. The system
therefore adopts a persistent loop, overlapping one work item's output store
with the next item's loads (Figure~\ref{fig:selfevo-dsa}), and ultimately obtains
a gain of \SelfEvoDsaSparseFwdGenOne{} over the preceding generation
(Table~\ref{tab:selfevo}). This case illustrates the role of reference updates
in self-evolution: retaining an existing artifact as a performance reference
while reopening optimization through renewed diagnosis and architecture
analysis, rather than fixing its structure as a constraint on subsequent runs.

\section{Related Work}
\label{sec:related}

\paragraph{Operator optimization and programming infrastructure.}
High-performance operator implementations require computation and data access
to be organized according to computational characteristics and target hardware.
FlashAttention uses tiling to reduce device-memory accesses in
attention~\cite{dao2022flashattention}, while FlashInfer provides customizable,
efficient attention implementations for LLM inference serving
~\cite{ye2025flashinfer}. Translating such designs into efficient code requires
appropriate programming abstractions and compiler support: CUTLASS provides
composable high-performance computational components, and Triton provides a
programming language and compiler for GPU kernels, enabling developers to
express and implement computation partitioning and data organization
~\cite{cutlass,tillet2019triton}. Beyond supporting developers in expressing
designs, another line of work explores automating design choices. For example,
Mirage jointly searches algebraic and scheduling transformations through a
unified representation spanning kernels, thread blocks, and
threads~\cite{wu2024mirage}.

\paragraph{LLM-based operator generation and optimization.}
Prior work improves operator generation through feedback-driven iteration,
evolutionary search, and model training. GEAK generates Triton kernels through
a feedback-driven reasoning loop~\cite{wang2025geak}. STARK and Astra combine
multi-agent collaboration with search, profiling, and planning
~\cite{dong2025stark,wei2025astra}; AlphaEvolve combines LLMs with automated
evaluation for evolutionary search~\cite{novikov2025alphaevolve}. These methods
focus on code iteration within a single run, without treating architectural
decisions as separately revisable from measurements or using accepted kernels
as references for a new round of architectural analysis.

Kevin incorporates multi-turn implementation and execution feedback into
reinforcement learning, while CUDA-L1 uses contrastive reinforcement learning
to improve CUDA optimization~\cite{baronio2025kevin,li2025cudal1}. Training
changes the model that emits code, not the revision of architectural
decisions during optimization.

EGG fixes the algorithmic structure before hardware tuning, without revisiting
it based on subsequent measurements~\cite{han2026egg}. CuBridge adapts expert
attention kernels to target operators through an executable intermediate
representation, retaining the expert implementation's structure
~\cite{ma2026cubridge}. AKO optimizes kernels through a multi-agent workflow,
with a human orchestrator redirecting stalled searches~\cite{luo2026ako}.
Structural choices or search redirections in these workflows are not explicit
architectural decision revisions triggered by measurements.

Another line of work retains prior experience and uses it to guide later
generation. KernelClaw stores reusable expert optimization knowledge in
long-term memory and uses short-term memory to avoid repeated backtracking
within a run, replacing the model's implicit heuristics with expert
knowledge~\cite{sun2026kernelskill}. KLineage progressively simplifies expert
implementations through validated steps, distilling each step into a skill with
explicit applicability conditions for subsequent code generation
~\cite{zhang2026klineage}. Both retain knowledge or skills
obtained from the expert side and use them to guide code generation. They do
not store hardware mechanisms validated by independent probes during
optimization as knowledge shared across operators, and later runs do not cite
such entries when setting architectural decisions.

\sys separates architectural design from code implementation, organizing
milestone plans around explicit architectural decisions and revising those
decisions in response to measurement feedback. When structural bottlenecks
limit performance, the system revisits computation partitioning, dataflow,
or pipeline organization rather than only tuning within a fixed architecture.
For knowledge reuse, \sys distills hardware mechanisms during optimization,
validates them through independent probes on the target hardware, and stores
them in a knowledge base shared across operators and input shapes to guide
architectural design and implementation planning in subsequent runs. Existing
implementations can also become new references: their architectures are
reassessed through measurement rather than carried forward by directly
inheriting their source code.

\section{Limitations and Future Work}
\label{sec:limitations}

\paragraph{Architecture exploration and local optima.}
\sys revisits existing architectures through explicit planning and
feedback-driven revision, but exploration is limited by the agents' ability to
generate candidates, and optimization may settle at local optima. Future work
will investigate planning strategies that broaden the architecture search space
and increase candidate diversity while preserving autonomous exploration and
generalization.

\paragraph{Cross-hardware design reuse.}
Hardware mechanisms in the knowledge base depend on specific architectures and
applicability conditions; transfer to other hardware requires adaptation and
validation. Future work will investigate mechanism abstraction, adaptation, and
validation for cross-hardware reuse while retaining target-specific optimization.

\section{Conclusion}
\label{sec:conclusion}

% Chinese mirror: zh/conclusion.tex (main_zh.pdf; update when this changes).

We presented \sys, a self-evolving agent for kernel optimization. \sys
separates architectural design from code implementation and revises work
partitioning, dataflow, and pipeline organization based on performance
feedback. Across runs, it uses validated kernels as new references
and accumulates hardware mechanisms validated by independent probes in a
shared knowledge base. Reference updates enable further architectural
reassessment, while knowledge reuse carries optimization experience across
runs and operators; together, they support continued self-evolution.

With self-evolution and knowledge reuse, \sys achieves a geometric-mean
speedup of \SpeedupRatioMean{}\(\times\) over reference implementations from
expert libraries and other kernel-generation systems across \NumShapes{}
configurations covering \NumOps{} operators on NVIDIA Hopper and Blackwell
GPUs. On Ascend, it outperforms the per-operator public best on the public
case set of all \AscendOpCount{} level-4 operators in the \cannbench{}
Season~\CannSeason{} final leaderboard, with a geometric-mean performance
gain of \AscendGeomeanLead{}. Public submissions generated by \sys rank first
on all three \cannbench{}~\CannVTwoVersion{} solution leaderboards
(\CannVTwoBoards{}) and on the SOL-ExecBench FlashInfer-Bench leaderboard
as of \LeaderboardSnapshot{}.

\bibliographystyle{ACM-Reference-Format}
\bibliography{references}

@misc{chen2026justgoal,
  author        = {Chen, Yaojian and Peng, Dazhi and Zhu, Zhui and Li, Leshan
                   and Sun, Haojun and Zheng, Yiyun and He, Qingfeng
                   and Li, Shangzhan and Deng, Xianglong and Shen, Yifei
                   and Xiang, Shengye and Li, Zhen and Gan, Lin
                   and Li, Yuxuan and Han, Xu and Liu, Zhiyuan},
  title         = {{JustGoal}: With Only a Goal, the {AI} Baseline Is Above the Expert Kernels},
  year          = {2026},
  eprint        = {2610.36434},
  archiveprefix = {arXiv}
}

@misc{cannbenchleaderboard2026,
  author       = {{CANNBench}},
  title        = {{CANNBench} Public Leaderboard},
  year         = {2026},
  howpublished = {\url{https://cannbench.com/leaderboard}},
  note         = {CANN-Bench 2.0 solution snapshot as of 2026-10-30: ForgeKernel
                  ranks first on the Ascend 910C, 950DT, and 950PR leaderboards.
                  Historical Season 1, version 1.1, A3 (Ascend 910C) result:
                  ForgeKernel: total score 8566.1110; standard/hidden scores
                  4362.7547/4203.3563, passed cases 1060/1060 and 4185/4240,
                  geometric-mean speedups 1.7591600842 and 1.8543285162.
                  Accessed 2026-10-30}
}

@misc{solexecbenchleaderboard2026,
  author       = {{NVIDIA Research}},
  title        = {{SOL-ExecBench}: {FlashInfer-Bench} Leaderboard},
  year         = {2026},
  howpublished = {\url{https://research.nvidia.com/benchmarks/sol-execbench/leaderboard/collection/4/B200}},
  note         = {B200 collection leaderboard, evaluation stack v1.1.
                  Submission 76892 (ForgeKernel): SOL Score 0.815204,
                  Fast1 647/660, solved 26/26, Avg Speedup 5.5403.
                  Accessed 2026-10-30}
}

@inproceedings{dao2022flashattention,
  author    = {Dao, Tri and Fu, Daniel Y. and Ermon, Stefano and Rudra, Atri and R{\'e}, Christopher},
  title     = {FlashAttention: Fast and Memory-Efficient Exact Attention with {IO}-Awareness},
  booktitle = {Advances in Neural Information Processing Systems (NeurIPS)},
  year      = {2022}
}

@misc{shah2024flashattention3,
  author        = {Shah, Jay and Bikshandi, Ganesh and Zhang, Ying and Thakkar, Vijay and Ramani, Pradeep and Dao, Tri},
  title         = {FlashAttention-3: Fast and Accurate Attention with Asynchrony and Low-precision},
  year          = {2024},
  eprint        = {2407.08608},
  archiveprefix = {arXiv}
}

@inproceedings{ye2025flashinfer,
  author    = {Ye, Zihao and Chen, Lequn and Lai, Ruihang and Lin, Wuwei and Zhang, Yineng and Wang, Stephanie and Chen, Tianqi and Kasikci, Baris and Grover, Vinod and Krishnamurthy, Arvind and Ceze, Luis},
  title     = {FlashInfer: Efficient and Customizable Attention Engine for {LLM} Inference Serving},
  booktitle = {Proceedings of Machine Learning and Systems (MLSys)},
  year      = {2025},
  note      = {Best Paper Award. arXiv:2501.01005}
}

@inproceedings{tillet2019triton,
  author    = {Tillet, Philippe and Kung, Hsiang-Tsung and Cox, David},
  title     = {Triton: An Intermediate Language and Compiler for Tiled Neural Network Computations},
  booktitle = {Proceedings of the 3rd ACM SIGPLAN International Workshop on Machine Learning and Programming Languages (MAPL)},
  year      = {2019},
  note      = {DOI: 10.1145/3315508.3329973}
}

@misc{cutlass,
  author       = {Thakkar, Vijay and Ramani, Pradeep and Cecka, Cris and Shivam, Aniket and Lu, Honghao and Yan, Ethan and Kosaian, Jack and Hoemmen, Mark and Wu, Haicheng and Kerr, Andrew and others},
  title        = {{CUTLASS}: {CUDA} Templates for Linear Algebra Subroutines},
  year         = {2026},
  howpublished = {\url{https://github.com/NVIDIA/cutlass}},
  note         = {Version 4.8; includes CuTe layout algebra and the Python CuTe-DSL. Accessed 2026-09}
}

@misc{h100whitepaper,
  author       = {{NVIDIA Corporation}},
  title        = {{NVIDIA} {H100} {Tensor Core} {GPU} Architecture Whitepaper},
  year         = {2022},
  howpublished = {\url{https://resources.nvidia.com/en-us-hopper-architecture/nvidia-h100-tensor-c}},
  note         = {Accessed 2026-09}
}

@misc{blackwellbrief,
  author       = {{NVIDIA Corporation}},
  title        = {{NVIDIA} {Blackwell} Architecture Technical Overview},
  year         = {2024},
  howpublished = {\url{https://resources.nvidia.com/en-us-blackwell-architecture}},
  note         = {Accessed 2026-09}
}

@misc{ptxisa,
  author       = {{NVIDIA Corporation}},
  title        = {Parallel Thread Execution {ISA}, Version 9.4},
  year         = {2026},
  howpublished = {\url{https://docs.nvidia.com/cuda/parallel-thread-execution/}},
  note         = {\texttt{wgmma.mma\_async}, \texttt{cp.async.bulk.tensor} (TMA), and \texttt{tcgen05} sections. Accessed 2026-09}
}

@misc{deepseek2024v2,
  author        = {{DeepSeek-AI}},
  title         = {{DeepSeek-V2}: A Strong, Economical, and Efficient Mixture-of-Experts Language Model},
  year          = {2024},
  eprint        = {2405.04434},
  archiveprefix = {arXiv}
}

@misc{deepseek2025v32exp,
  author       = {{DeepSeek-AI}},
  title        = {{DeepSeek-V3.2-Exp}: Boosting Long-Context Efficiency with {DeepSeek} Sparse Attention},
  year         = {2025},
  howpublished = {\url{https://github.com/deepseek-ai/DeepSeek-V3.2-Exp/blob/main/DeepSeek_V3_2.pdf}},
  note         = {Tech report, September 2025}
}

@misc{kimiteam2025kimilinear,
  author        = {{Kimi Team}},
  title         = {Kimi Linear: An Expressive, Efficient Attention Architecture},
  year          = {2025},
  eprint        = {2510.26692},
  archiveprefix = {arXiv},
  note          = {Introduces Kimi Delta Attention (KDA)}
}

@misc{yang2024gateddeltanet,
  author        = {Yang, Songlin and Kautz, Jan and Hatamizadeh, Ali},
  title         = {Gated Delta Networks: Improving {Mamba2} with Delta Rule},
  year          = {2024},
  eprint        = {2412.06464},
  archiveprefix = {arXiv},
  note          = {ICLR 2025}
}

@misc{cao2026ksearch,
  author        = {Cao, Shiyi and Mao, Ziming and Gonzalez, Joseph E. and Stoica, Ion},
  title         = {{K-Search}: {LLM} Kernel Generation via Co-Evolving Intrinsic World Model},
  year          = {2026},
  eprint        = {2602.19128},
  archiveprefix = {arXiv}
}

@misc{ye2026cake,
  author        = {Ye, Zihao and Huang, Yingyi and Jin, Hongyi and Hou, Bohan and Shao, Junru and Yu, Zhongming and Chen, Jinqi and Cowan, Meghan and Cao, Shiyi and Xing, Shanli and Chen, Hanfeng and Grover, Vinod and Chen, Tianqi and Ceze, Luis},
  title         = {{CAKE}: Compiler-Agent Co-Design for Frontier Kernel Evolution},
  year          = {2026},
  eprint        = {2608.12629},
  archiveprefix = {arXiv}
}

@misc{liu2026kda,
  author        = {Liu, Sihao and Zhu, Ligeng and others},
  title         = {Kernel Design Agents ({KDA}): An Agent-Centric Workflow for High-Performance {CUDA} Kernels},
  year          = {2026},
  howpublished  = {\url{https://github.com/NVlabs/kda}},
  note          = {NVIDIA and MIT HAN Lab; software release}
}

@misc{kdacontest2026,
  author       = {Zou, Dongyun and Dong, Yixin and Guo, Junxian and others},
  title        = {{KDA-v0.5} Kernels for the {MLSys} 2026 {FlashInfer} {AI} Kernel Generation Contest},
  year         = {2026},
  howpublished = {\url{https://github.com/mit-han-lab/mlsys2026-flashinfer-contest-solution}},
  note         = {Speedups over the contest winners re-measured by the authors under the flashinfer-bench CUPTI protocol, not by the contest evaluator. Accessed 2026-09}
}

@misc{novikov2025alphaevolve,
  author        = {Novikov, Alexander and V{\~u}, Ng{\^a}n and Eisenberger, Marvin and others},
  title         = {AlphaEvolve: A coding agent for scientific and algorithmic discovery},
  year          = {2025},
  eprint        = {2506.13131},
  archiveprefix = {arXiv}
}

@misc{baronio2025kevin,
  author        = {Baronio, Carlo and Marsella, Pietro and Pan, Ben and Guo, Simon and Alberti, Silas},
  title         = {Kevin: Multi-Turn {RL} for Generating {CUDA} Kernels},
  year          = {2025},
  eprint        = {2507.11948},
  archiveprefix = {arXiv}
}

@inproceedings{li2025cudal1,
  author    = {Li, Xiaoya and Wang, Albert and Wang, Guoyin and Li, Jiwei and Shum, Chris},
  title     = {{CUDA-L1}: Improving {CUDA} Optimization via Contrastive Reinforcement Learning},
  booktitle = {International Conference on Learning Representations (ICLR)},
  year      = {2026},
  note      = {arXiv:2507.14111}
}

@misc{wang2025geak,
  author        = {Wang, Jianghui and Joshi, Vinay and Majumder, Saptarshi and others},
  title         = {{GEAK}: Introducing {Triton} Kernel {AI} Agent \& Evaluation Benchmarks},
  year          = {2025},
  eprint        = {2507.23194},
  archiveprefix = {arXiv}
}

@misc{sun2026kernelskill,
  author        = {Sun, Qitong and Han, Jun and Li, Tianlin and others},
  title         = {KernelClaw: A Multi-Agent Framework for {GPU} Kernel Optimization with Dual-Memory Mechanisms},
  year          = {2026},
  eprint        = {2603.10085},
  archiveprefix = {arXiv}
}

@misc{ma2026cubridge,
  author        = {Ma, Xing and Zhou, Yangjie and Sun, Wu and others},
  title         = {CuBridge: An {LLM}-Based Framework for Understanding and Reconstructing High-Performance Attention Kernels},
  year          = {2026},
  eprint        = {2605.05023},
  archiveprefix = {arXiv},
  note          = {ACL 2026}
}

@misc{dong2025stark,
  author        = {Dong, Juncheng and Yang, Yang and Liu, Tao and others},
  title         = {{STARK}: Strategic Team of Agents for Refining Kernels},
  year          = {2025},
  eprint        = {2510.16996},
  archiveprefix = {arXiv}
}

@misc{han2026egg,
  author        = {Han, Yaochen and Fan, Ke and Jiang, Hongxu and others},
  title         = {{EGG}: An Expert-Guided Agent Framework for Kernel Generation},
  year          = {2026},
  eprint        = {2606.26758},
  archiveprefix = {arXiv}
}

@misc{chen2025nvidiadeepseek,
  author       = {Chen, Terry and Xu, Bing and Devleker, Kirthi},
  title        = {Automating {GPU} Kernel Generation with {DeepSeek-R1} and Inference Time Scaling},
  year         = {2025},
  howpublished = {NVIDIA Developer Blog},
  note         = {\url{https://developer.nvidia.com/blog/automating-gpu-kernel-generation-with-deepseek-r1-and-inference-time-scaling/}}
}

@inproceedings{wu2024mirage,
  author    = {Wu, Mengdi and Cheng, Xinhao and Liu, Shengyu and others},
  title     = {Mirage: A Multi-Level Superoptimizer for Tensor Programs},
  booktitle = {19th USENIX Symposium on Operating Systems Design and Implementation (OSDI)},
  year      = {2025},
  note      = {arXiv:2405.05751}
}

@misc{luo2026ako,
  author        = {Luo, Mao and Li, Hongbin and Lin, Feng and Yi, Hanling and Huang, Zhe},
  title         = {Agentic Kernel Optimization: Generating State-of-the-Art {GPU} Kernels Without Hand-Written {CUDA}},
  year          = {2026},
  eprint        = {2608.14560},
  archiveprefix = {arXiv}
}

@misc{zhang2026klineage,
  author        = {Zhang, Shuoming and Yu, Qiuchu and Zhang, Yangyu and others},
  title         = {{KLineage}: Recovering the Missing When of Kernel Optimization by Deoptimizing Experts},
  year          = {2026},
  eprint        = {2605.28213},
  archiveprefix = {arXiv}
}

@misc{wei2025astra,
  author        = {Wei, Anjiang and Sun, Tianran and Seenichamy, Yogesh and others},
  title         = {Astra: A Multi-Agent System for {GPU} Kernel Performance Optimization},
  year          = {2025},
  eprint        = {2509.07506},
  archiveprefix = {arXiv}
}

%% The appendix rides in this PDF; the body \ref's into it by section label.
% Bound into main.tex by the arXiv edition; the body \ref's here by section
% label (sec:app:judges / sec:app:refs / sec:app:shapes). The argument still
% stands without it: that source-specific tolerance tests, isolation,
% budgets, the /goal gate and which reference each cell is sealed against are
% all stated in the body's Experimental Setup. What follows is the per-judge,
% per-reference and per-shape detail that section names but does not carry.
% Keep zh/appendix.tex in sync.
\appendix
\section{Correctness Judges}
\label{sec:app:judges}

Correctness validation first follows the official or benchmark tolerance test
associated with each operator source. Operators covered by FlashInfer-Bench,
including those in the SOL-ExecBench FlashInfer-Bench collection, use
FlashInfer-Bench's tolerance tests; FlashAttention operators use the official
FlashAttention tolerance tests; other operators use the criterion associated
with their source. Within this source mapping, the concrete judges include
DeepGEMM's aggregate criterion for GEMMs, a relative max-abs bound for
attention gradients, and a scaled tolerance plus a cosine bound for the GDN
and KDA (Kimi Delta Attention~\cite{kimiteam2025kimilinear}) state kernels,
whose small outputs would pass a fixed absolute tolerance trivially. The GDN
judges further add correlated-key and long-memory inputs that catch
state-decay shortcuts.

\section{Reference Versions and Entry Points}
\label{sec:app:refs}

Library references are FlashAttention-4 (CuTe
DSL), cuDNN~9 with cuDNN-frontend~1.28, DeepGEMM, FlashMLA, FlashInfer and
FlashQLA, and are measured at these entry
points: FlashAttention-4's forward and \texttt{\_flash\_attn\_bwd},
FlashInfer's fused \texttt{chunk\_gated\_delta\_rule}, FlashQLA's TileLang
backward, DeepGEMM's \texttt{fp8\_fp4\_mqa\_logits}, FlashMLA's
\texttt{flash\_mla\_sparse\_fwd}, and cuDNN-frontend's CuTe DSL radix
top-$k$, \texttt{sparse\_attention\_backward}, \texttt{indexer\_backward}
(three kernels summed), and the \texttt{sparse\_attn} +
\texttt{sparse\_indexer} score-recompute composition.
\cannbench{}~\AscendBenchVersion{} supplies the Ascend baselines; its public
repository releases through v1.0.0 as of this writing.
% Transcribed from the reference column of data/plain_agent_sm90.csv (the
% body table shows the library only); keep the two in sync.

\section{Campaign Shapes}
\label{sec:app:shapes}

Table~\ref{tab:shapes} lists the problem shape of every task and optimization run of
the main paper.

\begin{table*}[tp]
  \centering
  \scriptsize
  \caption{Input shapes for the evaluated tasks and optimization runs.}
  \label{tab:shapes}
  % AUTO-GENERATED by scripts/gen_forge_tables.py -- edit data/*.csv instead.
\begin{tabular}{@{}ll>{\raggedright\arraybackslash}p{11cm}@{}}
\toprule
Operator & GPU & Shape \\
\midrule
\multicolumn{3}{@{}l}{\emph{\goalagent{} ceiling}} \\[1pt]
attn\_fwd\_gqa & H100 & B=2 S=4096 H\_Q=16/H\_KV=1 D=128 bf16 causal \\
fa\_bwd\_gqa & H100 & B=10 N=4096 H\_Q=16/H\_KV=2 D=64 fp16 causal \\
gdn\_prefill\_fwd & H100 & N=8192 h\_qk=16/h\_v=32 d=128 bf16 \\
gdn\_prefill\_bwd & H100 & N=8192 h\_qk=16/h\_v=32 d=128 bf16 \\
dsa\_indexer\_fwd & H100 & s\_q=4096 s\_kv=32768 H=64 D=128 fp8 \\
dsa\_indexer\_topk & H100 & logits fp32 [4096 x 32768] k=1024 \\
dsa\_sparse\_fwd & H100 & s\_q=4096 s\_kv=32768 h=128 d=512 topk=1024 \\
dsa\_sparse\_bwd & H100 & s\_q=32768 s\_kv=8192 h=128 d=512 topk=1024 \\
dsa\_score\_recompute & H100 & s\_q=32768 s\_kv=8192 topk=1024 \\
dsa\_indexer\_bwd & H100 & s\_q=32768 s\_kv=8192 h=64 d=128 topk=1024 \\
\midrule
\multicolumn{3}{@{}l}{\emph{\sys optimization runs}} \\[1pt]
attn\_fwd (GQA 16:1) & H100 & B=2 S=4096 H=16/1 D=128 bf16 causal \\
attn\_bwd (MHA) & H100 & B=10 N=4096 H=16 D=64 fp16 causal \\
attn\_bwd (GQA 8:1) & H100 & B=10 N=4096 H=16/2 D=64 fp16 causal \\
fp8\_gemm & H100 & M=N=K=4096 blk 128 \\
fp8\_grouped\_gemm & H100 & G=8 N=K=4096 blk 128 \\
gemm (fp16) & H100 & M=N=K=8192 fp16 \\
conv2d\_fwd & H100 & N=128 C=K=256 28x28 3x3 s1 \\
conv2d\_wgrad & H100 & N=128 C=K=256 28x28 3x3 s1 \\
conv2d\_dgrad & H100 & N=128 C=K=256 56x56 3x3 s2 \\
conv3d\_fwd & H100 & N=8 C=K=256 8x28x28 3x3x3 \\
dsa\_indexer\_fwd & H100 & s\_q=4096 s\_kv=32768 H=64 D=128 fp8 \\
dsa\_indexer\_topk & H100 & 4096x32768 fp32 k=1024 \\
dsa\_sparse\_fwd & H100 & s\_q=4096 s\_kv=32768 h=128 d=512 topk=1024 \\
dsa\_sparse\_bwd & H100 & s\_q=32768 s\_kv=8192 h=128 d=512 topk=1024 \\
dsa\_score\_recompute & H100 & s\_q=32768 s\_kv=8192 topk=1024 \\
dsa\_indexer\_bwd & H100 & s\_q=32768 s\_kv=8192 H=64 D=128 topk=1024 \\
gdn\_prefill\_fwd & H100 & N=8192 h=16/32 d=128 bf16 \\
gdn\_prefill\_bwd & H100 & N=8192 h=16/32 d=128 bf16 \\
gqa\_paged\_decode & H100 & workload 7a7fc288 (61535 pages) \\
mla\_paged\_decode & H100 & workload 5bef8d88 (B=64; 75145 tokens) \\
mla\_paged\_prefill & H100 & workload fe63f292 (B=18; 15883 q) \\
trimul & H100 & S=1024 D=384 H=128 (case 6) \\
attn\_bwd (MHA) & B200 & B=8 N=4096 H=16 D=128 bf16 causal \\
fp4\_grouped\_gemm & B200 & G=8 M=4096/group N=K=4096 \\
sparse\_mla\_prefill & B200 & s\_q=4096 s\_kv=32768 H\_Q=128 H\_KV=1 D=512 topk=1024 bf16 attn\_sink \\
fused\_attn\_rope\_cast & B200 & s\_q=4096 s\_kv=32768 H\_Q=128 H\_KV=1 D=512 rope=64 topk=1024 n\_wv\_group=16 bf16->fp8 attn\_sink q\_norm \\
dsa\_sparse\_attn\_paged & B200 & workload 564007ac (T=8; 8462 pages) \\
gdn\_prefill\_varlen & B200 & workload a9540651 (T=8192 N=20) \\
recurrent\_kda\_prefill & B200 & T=8192 H=96 K=V=128 bf16 \\
\bottomrule
\end{tabular}

\end{table*}

\section{Per-Task Public-Benchmark Speedups}
\label{sec:app:benchmark-speedups}

Tables~\ref{tab:app:cann-speedups} and~\ref{tab:app:sol-speedups} list all
\CannLeaderboardOps{} CANN-Bench Season~\CannSeason{} A3 operators and all
\SolExecProblems{} SOL-ExecBench FlashInfer-Bench kernels, respectively.
All values are speedup ratios against the corresponding benchmark baseline;
values below one indicate slowdowns and are retained.

For CANN-Bench, we report the standard-set, hidden-set, and combined
speedups from the Season~\CannSeason{} final results, checked on
\CannSeasonSnapshot{}. The combined column uses the benchmark's weighted
geometric mean with standard/hidden weights \(0.2/0.8\).
Reported ratios are rounded to three decimals; they do not imply that all
hidden cases passed. GRU and LSTM use hardware-bound-derived baselines
rather than reference implementations~\cite{cannbenchleaderboard2026}.

For SOL-ExecBench, we use ForgeKernel's per-kernel leaderboard entries on
B200 with evaluation stack v1.1, accessed on \LeaderboardSnapshot{}.
Each ratio is the geometric-mean speedup across that kernel's workloads
against the official \emph{Scoring Baseline}, not the
\emph{Reference Implementation} or \emph{SOL Bound}.
The geometric mean across all \SolExecProblems{} kernel ratios is
\(\SolExecAvgSpeedup\times\)~\cite{solexecbenchleaderboard2026}.

\begin{table*}[tp]
  \centering
  \scriptsize
  \setlength{\tabcolsep}{3pt}
  \caption{Per-operator CANN-Bench Season~\CannSeason{} A3 speedups
    (\(\times\)) against the benchmark baseline on Ascend \CannLeaderboardHw{}.}
  \label{tab:app:cann-speedups}
  \begin{tabular*}{\textwidth}{@{\extracolsep{\fill}}lcrrr@{\hspace{14pt}}lcrrr@{}}
    \toprule
    Operator & Level & Standard & Hidden & Combined &
    Operator & Level & Standard & Hidden & Combined \\
    \midrule
    % AUTO-GENERATED by scripts/gen_benchmark_speedups.py; edit data/*_benchmark_speedups.csv.
Gelu & L1 & 0.910 & 3.263 & 2.527 & DepthwiseConv2D & L3 & 1.905 & 1.143 & 1.266 \\
SwiGlu & L1 & 2.113 & 4.013 & 3.530 & TopK & L3 & 1.286 & 0.855 & 0.927 \\
ForeachNorm & L1 & 2.088 & 3.707 & 3.305 & MoeFinalizeRouting & L3 & 2.306 & 3.229 & 3.019 \\
Sigmoid & L1 & 0.972 & 4.679 & 3.417 & Dilation2D & L3 & 2.086 & 1.190 & 1.331 \\
Exp & L1 & 2.717 & 4.624 & 4.157 & MhcSinkhorn & L3 & 3.188 & 2.603 & 2.711 \\
MaskedScale & L1 & 2.652 & 4.837 & 4.289 & StridedSlice & L3 & 3.667 & 3.958 & 3.898 \\
Mish & L1 & 2.369 & 3.965 & 3.577 & DequantSwigluQuant & L3 & 2.512 & 3.744 & 3.457 \\
ForeachAddcdivScalar & L1 & 1.804 & 4.997 & 4.076 & WeightQuantBatchMatmul & L3 & 1.299 & 2.143 & 1.939 \\
GridSampler3D & L2 & 1.343 & 0.861 & 0.941 & EngramGateFusion & L3 & 0.988 & 0.968 & 0.972 \\
Cummin & L2 & 1.285 & 1.032 & 1.078 & Conv3DBackpropFilter & L3 & 1.758 & 1.104 & 1.212 \\
Gcd & L2 & 5.704 & 3.296 & 3.678 & Unique & L3 & 1.487 & 0.589 & 0.709 \\
DynamicQuant & L2 & 2.270 & 5.754 & 4.777 & GroupedMatmul & L3 & 1.868 & 2.406 & 2.287 \\
CrossEntropyLoss & L2 & 2.253 & 1.691 & 1.791 & Conv2D & L3 & 1.019 & 1.078 & 1.066 \\
ApplyRotaryPosEmb & L2 & 2.510 & 2.216 & 2.272 & AdaptiveAvgPool3D & L3 & 3.478 & 3.074 & 3.151 \\
GroupNorm & L2 & 1.589 & 2.132 & 2.011 & MoeReRouting & L3 & 2.015 & 3.201 & 2.918 \\
Softmax & L2 & 1.840 & 2.380 & 2.261 & Transpose & L3 & 1.213 & 1.913 & 1.746 \\
RmsNorm & L2 & 1.709 & 3.032 & 2.703 & ROIAlign & L3 & 0.842 & 0.631 & 0.669 \\
Scatter & L2 & 1.028 & 0.363 & 0.447 & QuantMatmul & L3 & 1.146 & 1.808 & 1.651 \\
Gather & L2 & 2.512 & 2.407 & 2.428 & GroupedMatmulSwigluQuant & L4 & 2.418 & 1.967 & 2.050 \\
UnsortedSegmentSum & L2 & 1.585 & 1.669 & 1.652 & LSTM & L4 & 0.450 & 0.147 & 0.184 \\
ResizeBilinear & L2 & 5.962 & 10.397 & 9.302 & MLA & L4 & 1.139 & 1.649 & 1.532 \\
ArgMax & L2 & 2.921 & 2.926 & 2.925 & GRU & L4 & 0.546 & 0.218 & 0.262 \\
ApplyAdamW & L2 & 1.965 & 4.378 & 3.730 & MlaProlog & L4 & 1.562 & 1.198 & 1.263 \\
Maximum & L2 & 1.570 & 4.117 & 3.395 & MHA & L4 & 1.198 & 1.112 & 1.128 \\
NMS & L3 & 3.191 & 0.981 & 1.242 & SparseFlashAttention & L4 & 1.461 & 0.207 & 0.306 \\
MoeGatingTopKSoftmax & L3 & 1.931 & 1.478 & 1.559 & GQA & L4 & 1.184 & 1.423 & 1.371 \\
AddRmsNormDynamicQuant & L3 & 2.459 & 3.250 & 3.074 &  &  &  &  & 
 \\
    \bottomrule
  \end{tabular*}
\end{table*}

\begin{table*}[tp]
  \centering
  \scriptsize
  \caption{Per-kernel SOL-ExecBench FlashInfer-Bench speedups against
    Scoring Baseline on B200 (evaluation stack v1.1).}
  \label{tab:app:sol-speedups}
  \begin{tabular*}{\textwidth}{@{\extracolsep{\fill}}lr@{}}
    \toprule
    Kernel & Speedup (\(\times\)) \\
    \midrule
    % AUTO-GENERATED by scripts/gen_benchmark_speedups.py; edit data/*_benchmark_speedups.csv.
\texttt{001\_fused\_add\_rmsnorm\_h2048} & 1.4651 \\
\texttt{002\_fused\_add\_rmsnorm\_h4096} & 4.4455 \\
\texttt{003\_fused\_add\_rmsnorm\_h7168} & 1.8506 \\
\texttt{004\_gemm\_n128\_k2048} & 2.4275 \\
\texttt{005\_gemm\_n256\_k7168} & 2.3825 \\
\texttt{006\_gemm\_n2048\_k4096} & 1.4514 \\
\texttt{007\_gemm\_n4096\_k4096} & 1.3595 \\
\texttt{008\_gemm\_n4096\_k14336} & 1.5171 \\
\texttt{009\_gemm\_n5120\_k2048} & 1.1265 \\
\texttt{010\_gemm\_n6144\_k4096} & 1.2387 \\
\texttt{011\_gemm\_n28672\_k4096} & 1.3481 \\
\texttt{012\_gqa\_paged\_decode\_h32\_kv4\_d128\_ps1} & 16.1650 \\
\texttt{013\_gqa\_paged\_decode\_h32\_kv8\_d128\_ps1} & 9.9521 \\
\texttt{014\_gqa\_paged\_prefill\_causal\_h32\_kv4\_d128\_ps1} & 159.3385 \\
\texttt{015\_gqa\_paged\_prefill\_causal\_h32\_kv8\_d128\_ps1} & 133.7091 \\
\texttt{016\_gqa\_ragged\_prefill\_causal\_h32\_kv4\_d128} & 17.5878 \\
\texttt{017\_gqa\_ragged\_prefill\_causal\_h32\_kv8\_d128} & 16.0272 \\
\texttt{018\_mla\_paged\_decode\_h16\_ckv512\_kpe64\_ps1} & 31.9883 \\
\texttt{019\_mla\_paged\_prefill\_causal\_h16\_ckv512\_kpe64\_ps1} & 93.2019 \\
\texttt{020\_moe\_fp8\_block\_scale\_ds\_routing\_topk8\_ng8\_kg4\_e32\_h7168\_i2048} & 28.1198 \\
\texttt{021\_rmsnorm\_h128} & 2.3360 \\
\texttt{022\_rmsnorm\_h512} & 2.3813 \\
\texttt{023\_rmsnorm\_h1536} & 4.8512 \\
\texttt{024\_rmsnorm\_h2048} & 2.4907 \\
\texttt{025\_rmsnorm\_h4096} & 2.7274 \\
\texttt{026\_rmsnorm\_h7168} & 3.6834
 \\
    \bottomrule
  \end{tabular*}
\end{table*}

\section{Knowledge-Reuse Provenance}
\label{sec:app:kb-provenance}

\begin{figure*}[tp]
  \centering
  \includegraphics[width=\textwidth]{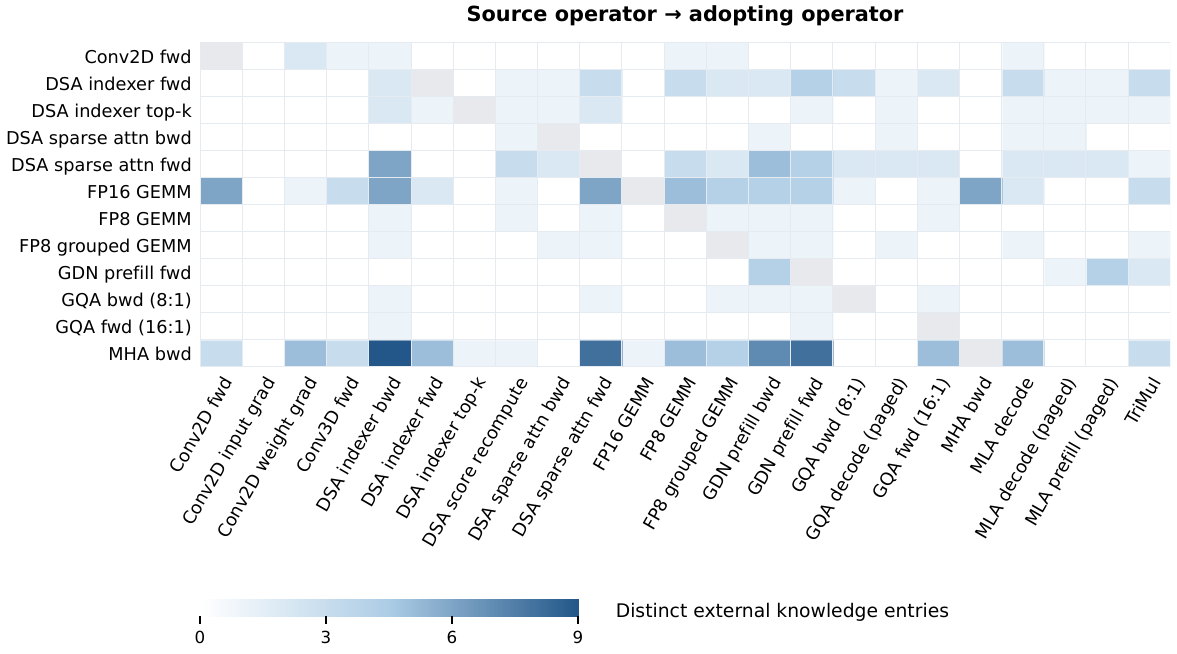}
  \caption{Sources and recipients of reused hardware knowledge.}
  \label{fig:kb-source-matrix}
\end{figure*}

Figure~\ref{fig:kb-source-matrix} aggregates the SM90 run snapshot by source
and adopting operator configuration. Each cell counts distinct external entries;
repeated use across runs is deduplicated. White cells have no qualifying record,
and gray cells denote self-reuse, which is excluded. All 23 recipient configurations
are retained, including the zero-adoption configuration; source rows without any
qualifying outgoing reuse are omitted. The cohort follows the original provenance
plot: MHA and GQA attention variants are separated, and MHA forward is not a
recipient because it is outside that plot's evaluated configuration set.
This is a provenance cohort, not a census of every evaluated task.

Entry origins come from recorded birth cells, falling back to the earliest run
of the originating operator. The original run-group ordering and temporal
exclusions are retained. Adoption uses the milestone-citation and decision-delivery
snapshot, rather than mere retrieval. Where commit lineage is available, delivery
is checked against the best correct implementation; otherwise the extractor falls
back to process records. Thus the snapshot does not independently certify final-code
retention for every entry, and zero denotes no qualifying recorded adoption.
The two examples in Figure~\ref{fig:kb-provenance}(b) show selected entries from
the recorded exchanges, not an exhaustive list or a causal performance attribution.

\end{document}